\documentstyle[twoside,amssymb,11pt]{article}

\newcommand{\nc}{\newcommand}
\nc{\la}{\lambda} \nc{\alf}{\alpha}
\nc{\tht}{\theta}  \nc{\be}{\beta}  \nc{\eps}{\epsilon} \nc{\ze}{\zeta}
\nc{\ga}{\gamma}  \nc{\De}{\Delta}  \nc{\Ga}{\Gamma}  \nc{\La}{\Lambda}
\nc{\de}{\delta} \nc{\si}{\sigma}  \nc{\ka}{\kappa}   \nc{\Si}{\Sigma}
\nc{\om}{\omega}  \nc{\qq}{\quad\quad}                \nc{\Om}{\Omega}
\nc{\nf}{\infty}   \nc{\dl}{\mathop{\smash{\cal L}}}  \nc{\black}{\rule{3mm}{3mm}}
\nc{\ra}{\rightarrow}  \nc{\ol}{\overline}  \nc{\und}{\underline}
\nc{\beq}{\begin{equation}}  \nc{\pt}{\partial}  \nc{\nin}{\noindent}
\nc{\eeq}{\end{equation}}
\nc{\beqa}{\begin{eqnarray}}  \nc{\dst}{\displaystyle}
\nc{\eeqa}{\end{eqnarray}} \nc{\nnb}{\nonumber}
\nc{\bs}{\backslash}        \nc{\mb}{\mathbb}

\newcounter{muni}
\newenvironment{remunerate}{\begin{list}{{\rm \arabic{muni}.}}
{\usecounter{muni}
\setlength{\leftmargin}{0pt}\setlength{\itemindent}{38pt}}}{\end{list}}

\nc{\brm}{\begin{remunerate}}   \nc{\erm}{\end{remunerate}}

\newtheorem{nlem}{Lemma} \newtheorem{nth}{Theorem}
\nc{\stg}{\mathop{\smash{*}}}
\nc{\st}{\mathop{\smash{\delta}}}
\nc{\barr}{\begin{array}}   \nc{\earr}{\end{array}}   \nc{\dg}{\dagger}
\nc{\mtvb}{\mathversion{bold}}   \nc{\mtvn}{\mathversion{normal}}
\nc{\ti}{\tilde}  \nc{\wti}{\widetilde} \nc{\wh}{\widehat}

\begin{document}

\begin{titlepage}
\begin{flushright}
April 2013
\end{flushright} 
\vskip 2.0truecm
\centerline{\large \bf ON A CLASS OF INTEGRABLE SYSTEMS}
\vskip 0.3truecm
\centerline{\large \bf WITH A QUARTIC FIRST INTEGRAL}
\vskip 1.0truecm 
\centerline{\bf Galliano VALENT${\ }^{*}\ \dagger$}
\vskip 1.0truecm 

\centerline{${}^{*}$\it Laboratoire de Physique Th\'eorique et des
Hautes Energies}
\centerline{\it Unit\'e associ\'ee au CNRS UMR 7589}
\centerline{\it 2 Place Jussieu, 75251 Paris Cedex 05, France} 
\vskip 1truecm

\centerline{${}^{\dagger}$\it Aix-Marseille Universit\'e et Universit\'e de Toulon}
\centerline{\it Centre de Physique Th\'eorique}
\centerline{\it Unit\'e associ\'ee au CNRS UMR 7332}
\centerline{\it Case 907, 13288 Marseille Cedex 9, France} 
\vskip 3truecm

\begin{abstract} We generalize, to some extent, the results on integrable geodesic flows on two dimensional  manifolds with a quartic first integral in the framework laid down by Selivanova and Hadeler. The local structure is first determined by a direct integration of the differential system 
which expresses the conservation of the quartic observable and is seen to involve a finite number of parameters. The global structure is studied in some details and leads to a class of models living 
on the manifolds ${\mb S}^2,{\mb H}^2$ or ${\mb R}^2$. As special cases we recover 
Kovalevskaya's integrable system and a generalization of it due to Goryachev.
\end{abstract}
\end{titlepage}

\nc{\pth}{P_{\tht}}  \nc{\pf}{P_{\phi}}  \nc{\pr}{P_{\rho}} \nc{\px}{P_x} \nc{\pz}{P_{\ze}}
\nc{\pF}{P_{\Phi}}

\section{Introduction}
In 1999 Selivanova \cite{Se2} has studied a class of integrable models, in two dimensional  manifolds, with a quartic first integral which generalized Kovalevskaya's system. In a 
collaboration with Hadeler \cite{hs} the explicit local structure of these models was given and 
led to new globally defined systems on ${\mb S}^2$. The aim of this article is to study a  generalization of these models and to determine which manifolds are involved. 

The plan of the article is the following: in Section 2 we describe the general 
setting of our integrable models and solve the differential system giving their local structure. 
This requires to split the analysis in two cases, according to whether a parameter $\mu$ 
vanishes or not.

In Section 3 we analyze, in the case where $\mu$ vanishes, the global structure of the systems and the nature of the manifolds.

In Section 4 we consider the case where $\mu$ does not vanish. The integrable systems of Kovalevskaya and its generalization due to Goryachev appear in this class.

In Section 5 we prove that all of these models do not exhibit integrals of degree 
less or equal to three in the momenta.

In Section 6 some conclusions are presented, followed by Appendix A, devoted to a summary of definitions and formulas used throughout the article.

\section{Local structure}
Let us first present the general structure of the integrable systems to be dealt with.
 
\subsection{The setting}
We will start from the usual hamiltonian
\[H=K+V\]
where the kinetic energy and the potential are 
\beq\label{ham}
K=\frac 12\Big(\pth^2+a(\tht)\,\pf^2\Big),\qq V=f(\tht)\,\cos\phi+g(\tht)\qq\qq f\not\equiv 0\eeq
while the quartic integral will have the form $\ Q=Q_4+Q_2+Q_0\ $ where
\beq\label{obsQ}\barr{l}
Q_4=\la\,\pf^4+2\mu\,H\,\pf^2\qq\qq\qq (\la,\,\mu)\in\,{\mb R}^2\bs\,\{(0,0)\},\\[4mm]
Q_2=\alf\,\cos\phi\,\pth^2+2\be(\tht)\,\sin\phi\,\pth\,\pf
+(\ga_0+\ga(\tht)\,\cos\phi)\pf^2\\[4mm]
Q_0=q(\tht,\phi).\earr\eeq
The parameters $\,\alf$ and $\,\ga_0$ are free and let us add the following comments:
\brm
\item The flow of $H$ describes the geodetic motion for the metric 
\beq\label{metricG}
g=d\tht^2+\frac{d\phi^2}{a(\tht)}.\eeq
We will consider only {\em riemannian} metrics.
\item The restriction $f\not\equiv 0$ is quite essential to obtain a truly quartic first integral. 
Indeed if $f$ vanishes $\pf$ is conserved and $Q_4$ becomes reducible.
\item If one takes $\ga_0(\tht)$ instead of a constant then its derivative must vanish 
for $\,Q$ to be an integral. So we take $\ga_0$ to be a constant.
\item If in $Q_2$ one takes $(\alf_0(\tht)+\alf(\tht)\,\cos\phi)\,\pth^2$, the two 
functions $\alf_0$ and $\alf$ must be constants for $\,Q$ to be an integral. Using the relation
\[\alf_0\,\pth^2=\alf_0\,(2H-a(\tht)\,\pf^2-V)\]
we see that the piece involving the hamiltonian  is reducible while the term $\,\pf^2$ can be 
included in $\ga_0$ and the term $\,V$ can be included in $q(\tht,\phi)$. Hence we have set $\alf_0=0$. 
\item If in $Q_2$ one adds a term $\,2\be_0(\tht)\,\pth\,\pf$, with a non-vanishing $\be_0$, 
then  $f$ must vanish for $\,Q$ to be an integral. So we take $\,\be_0(\tht)\equiv 0$.
\item Let us describe in this setting Kovalevskaya integrable system which is globally defined 
on ${\mb S}^2$. Using the $so(3)$ generators defined in Appendix A it is given by
\beq
H=\frac 12(L_1^2+L_2^2+2\,L_3^2)+k\,x\qq
Q=\left|\frac 12(L_1+iL_2)^2-k(x+iy)\right|^2,\eeq
easily transformed into 
\beq\label{KovaH}
2H=\pth^2+\frac{1+\sin^2\tht}{\sin^2\tht}\,\pf^2+2k\,\sin\tht\,\cos\phi
\eeq
which describes the geodesic flow on the metric
\beq\label{Kovamet}
g=d\tht^2+\frac{\sin^2\tht}{1+\sin^2\tht}\,d\phi^2.
\eeq
Its quartic first integral may be simplified into $\,\wh{Q}\equiv Q-H^2$ giving
\beq\label{KovaQ}
\wh{Q}=\pf^4-2H\,\pf^2-2k\,\cos\tht\,\left(\sin\phi\,\pth
+\frac{\cos\phi}{\tan\tht}\,\,\pf\right)\pf+k^2\,\sin^2\tht\,\sin^2\phi
\eeq
which does fit with (\ref{obsQ}) for $\alf=0$. 
\erm

\subsection{The differential system}
It appears convenient to define, instead of $\tht$, a new variable $t$ 
such that $\,dt=a(\tht)d\tht$. The integrable system becomes
\beq\left\{\barr{l}\dst 
H=\frac 12(a^2\,P_t^2+a\,\pf^2)+f\,\cos\phi+g\\[4mm]\dst 
Q=\la\,\pf^4+2\mu\,H\,\pf^2+\Big(\alf a^2\,\cos\phi\,P_t^2+2\be a\,\sin\phi\,P_t\,\pf+(\ga_0+\ga\,\cos\phi)\pf^2\Big)+q.\earr\right.\eeq
Let us prove:

\begin{nth} The constraint $\,\{H,Q\}=0$ is equivalent to the differential system
\beq\label{dsystem}\barr{ll}\dst
(a)\qq & \dst \dot{\be}=\frac{\alf}{2}-\mu\,\frac fa\qq \dot{\ga}=-2\be+\alf\,\dot{a}\qq 
\ga+\be\,\dot{a}=4\la\,\frac fa+2\mu\,f \\[4mm]\dst 
(b)\qq & \dst \dot{\De}+2\alf\,\dot{f}=-2\be\,\frac fa\qq\qq  
\be\,\dot{g}+\alf\,g-(\ga_0+2\mu\,g)\frac fa=L,\earr\eeq
where $L$ is some constant and
\[\Delta=\be\,\dot{f}-(\ga+2\mu\,f)\frac fa.\]
\end{nth}

\nin{\bf Proof:} In the Poisson bracket $\,\{H,Q\}$ the terms which would be of degree 5 in the momenta  vanish identically. The terms of degree 3 give the equations (a) and the last piece 
of degree 1 yields 
\[\barr{l}\dst 
\pt_{\phi}\Big(\frac q2\Big)=\Big(\be\,\dot{g}-(\ga_0+2\mu\,g)\,\frac fa\Big)\sin\phi
+\Delta\,\sin\phi\,\cos\phi\qq \\[4mm]\dst 
\pt_t\Big(\frac q2\Big)=\alf\,\dot{f}+\alf\,\dot{g}\,\cos\phi
-\Big(\alf\,\dot{f}+\be\,\frac fa\Big)\sin^2\phi.\earr\]
These partial differential equations give the two integrability conditions (b) in (\ref{dsystem}). 
When these relations hold we get
\beq\label{solq}
q=2\alf\,f+2(\alf\,g-L)\cos\phi+\De\,\sin^2\phi,\eeq 
which concludes the proof. $\quad\Box$ 

Let us now integrate this differential system, splitting the analysis in two cases.

\subsection{First case: $\mu=0$ }
Let us begin with:

\begin{nth} The local structure of the integrable system is
\beq\label{fcaseH} 
2\,H=\frac 1{\be^2}\Big(F\,P_x^2+\sqrt{F}\,\pf^2+\ka\,\sqrt{F}\,G'\,\cos\phi
+l\,G+m\,x+n\Big)\eeq
with
\[\be^2=b_0+\alf\,x\qq F=x^4+c_2\,x^2+c_1\,x+c_0\qq  G=\sqrt{F}-x^2-\frac{c_2}{2}\]
and for the quartic integral
\beq\label{fcaseQ}\barr{l}\dst 
Q=\pf^4+2\ka\Big(\alf\frac{F}{\be^2}\,\cos\phi\,P_x^2+2\,\sqrt{F}\,\sin\phi\,P_x\,\pf
+(\alf\frac{\sqrt{F}}{\be^2}-2x)\,\cos\phi\,\pf^2\Big)+2l\,\pf^2\\[4mm]\dst  
\hspace{1cm}+2\ka^2\left(\alf\,\frac{\sqrt{F}G'}{\be^2}\,\cos^2\phi-G\,\sin^2\phi\right)
+2\ka\Big(\alf\,\frac{h}{\be^2}-m\Big)\cos\phi.\earr\eeq
where
\[h=l\,G+m\,x+n\]
and all the constants are real. 
\end{nth}

\nin{\bf Proof:}
The differential system (\ref{dsystem}) reduces to
\beq\label{ds1}\barr{l}\dst 
\dot{\be}=\frac{\alf}{2}\qq\qq\dot{\ga}=-2\be+\alf\,\dot{a}\qq\qq 
\ga+\be\,\dot{a}=4\la\,\frac fa\qq \la\neq 0\\[4mm]\dst 
D_t\left(\be\,\dot{f}-\ga\,\frac fa+2\alf\,f\right)=-2\be\,\frac fa\qq 
\be\,\dot{g}+\alf\,g=L+\ga_0\,\frac fa.\earr
\eeq
We will define a new variable 
$\,2x=\alf\,a-\ga$ for which we have $\,\dot{x}=\be$ and we will demand that $\be\not\equiv 0$. Denoting by a prime a derivative 
with respect to $x$ we get from the first and the third relations in (\ref{ds1}) 
\beq\label{fcts1}
\be^2=b_0+\alf\,x\qq\qq 4\la\,\frac fa=(\be^2\,a)'-2x.\eeq
The fourth relation in (\ref{ds1}), using the variable $x$, can be integrated once and gives 
\beq\label{Delta1}
4\la\left(\be^2\,f'+(\alf\,a+2x)\,\frac fa\right)=c_2+2x^2-2\be^2\,a\eeq
where $\,c_2$ is a constant. Getting rid of $f$ 
using the second relation in (\ref{fcts1}), gives an ODE for $\,a$ which is
\beq
\Big(\be^4\,a^2\Big)''=12x^2+2c_2\qq\Longrightarrow\qq a=\frac{\sqrt{F}}{\be^2}\qq\qq  
F=x^4+c_2\,x^2+c_1\,x+c_0,\eeq
where we took the positive root for $a$ to secure the euclidean signature.

Upon the changes $\ga_0\to l$ and $L\to m/2$ we obtain 
\beq
f=\frac 1{4\la}\,\frac{\sqrt{F}\,G'}{\be^2}\qq\qq g=\frac 1{2\be^2}\left(mx+n+\frac l{2\la}\,G\right)
\qq G=\sqrt{F}-x^2-\frac{c_2}{2},\eeq
and $q$ follows from (\ref{solq}). Setting $\dst\ka=\frac 1{2\la}$ and a 
few scalings give for the quartic integral the formula (\ref{fcaseQ}). $\quad\Box$

\subsection{Second case: $\mu\neq 0$ }
In this case we have
\begin{nth}
The local form of the integrable system is
\beq\label{scaseH}
2\,H=\frac 1{\be^2}\Big(F\,P_x^2+\sqrt{F}\,\pf^2+\ka\,\sqrt{F}\,G'\,\cos\phi+l\,G+mx+n \Big)
\eeq
with
\beq\label{FB}
\left\{\barr{l}F=(x^2+d)^2-4\,r\,p(x)\qq p(x)=\alf\,x^3+(c_2-r\,\alf^2)\,x^2+c_1\,x+c_0\\[4mm]\dst 
\be^2=\frac{2\,p(x)}{x^2+d+\sqrt{F}}\qq\qq G=c_2+\alf\,x-\be^2\earr\right.
\eeq
and for the quartic integral
\beq\label{scaseQ}\barr{l}\dst
Q=r\,\pf^4+2\,H\,\pf^2\\[4mm]\dst 
\qq +\ka\left(\alf\frac F{\be^2}\,\cos\phi\,P_x^2+2\,\sqrt{F}\,\sin\phi\,P_x\,\pf
+(\alf\frac{\sqrt{F}}{\be^2}+2r\,\alf-2x)\cos\phi\,\pf^2\right)+l\,\pf^2\\[4mm]\dst 
\qq +\ka^2\left(\alf\frac{\sqrt{F}\,G'}{\be^2}\,\cos^2\phi-G\,\sin^2\phi\right)+\ka\left(\alf\frac h{\be^2}-m\right)\cos\phi.\earr\eeq
where
\[h=l\,G+m\,x+n\]
and all the constants are real.
\end{nth}

\nin{\bf Proof:}
The differential system (\ref{dsystem}) becomes 
\beq\label{ds2}\barr{l}\dst 
\dot{\be}=\frac{\alf}{2}-\mu\,\frac fa\qq\qq\dot{\ga}=-2\be+\alf\,\dot{a},\qq\qq 
\ga+\be\,\dot{a}=4\la\,\frac fa+2\mu\,f\\[4mm]\dst 
D_t\left(\be\,\dot{f}-(\ga+2\mu f)\,\frac fa+2\alf\,f\right)=-2\be\,\frac fa\qq 
\be\,\dot{g}+a\alf\,g-(\ga_0+2\mu g)\,\frac fa=L.\earr
\eeq
where $L$ is an integration constant.

Let us define a new variable 
\beq
2x=\alf\,(a+2\,r)-\ga\qq\qq\dot{x}=\be\neq 0\qq\qq r=\frac{\la}{\mu}\in{\mb R}.
\eeq
Denoting by a prime a derivative with respect to the new variable $x$, 
the first and the third relations in (\ref{ds2}) become
\beq\label{fa2}
2\mu\,\frac fa=\alf-(\be^2)'\qq\qq 2\,r\,(\be^2)'+(a\,\be^2)'=2x\qq\qq
\eeq
and the last one implies
\beq\label{fcta2}
a=\frac{x^2+d-2\,r\,\be^2}{\be^2}\equiv\frac{\sqrt{F}}{\be^2}\qq\qq F=(x^2+d-2\,r\,\be^2)^2
\eeq
where $d$ is an integration constant.

The fourth relation in (\ref{ds2}) can be written
\beq
D_t(\De+2\alf\,f)=-2\be\,\frac fa\qq\qq \De=\be\,\dot{f}-(\ga+2\mu\,f)\,\frac fa,\eeq
and when expressed in terms of the variable $x$ it integrates up to
\beq
\be^2\,(2\mu f)'+2\mu\,\frac fa\left(2x-2\,r\alf+a\,(\be^2)'\right)-2\be^2+2\alf x+2c_2=0\eeq
with a new integration constant $c_2$. Using in this preceding relation the formulas (\ref{fa2}) 
and (\ref{fcta2}), after some computations, one gets a simple ODE for $\,\be^2$: 
\beq
\Big(r\,(\be^2)^2-(x^2+d)\,\be^2\Big)''+6\alf x+2(c_2-\rho\alf^2)=0\eeq
which is readily integrated to
\beq
r\,(\be^2)^2-(x^2+d)\,\be^2+p(x)=0\qq\qq p(x)=\alf\,x^3+(c_2-r\,\alf^2)\,x^2+c_1\,x+c_0.
\eeq
Inserting this result into (\ref{fcta2}) gives $F$ and solving for $\be^2$ we get the 
the relations given in (\ref{FB}).
 
One obtains for $f$ and $g$ 
\[2f=\ka\frac{\sqrt{F}G'}{\be^2}\qq\qq 2g=\frac{h}{\be^2}\qq\qq h=l\,G+mx+n\]
and one gets $q$ using relation (\ref{solq}). After a few scalings one obtains the formula 
(\ref{scaseQ}) for $Q$ . $\quad\Box$

\vspace{2mm}
\nin {\bf Remarks:}
\brm
\item The structure of this integrable model is therefore described by a finite number of parameters, playing different roles. Firstly we have the parameters which define the metric: 
$\alf,\,b_0,\,c_0,\,c_1,\,c_2$ in the first case and $\alf,\,r,\,\,c_0,\,c_1,\,c_2,\,d$ 
in the second case; secondly we have the principal parameter $\ka$ which describes the bulk 
of the integrable system and thirdly we have secondary parameters $l,\,m,\,n$ which creep in 
through $g$, as defined in formula (\ref{ham}), and are of minor interest. 

\item Two integrable systems, with a quartic first integral, were derived in \cite{Se1} and 
in \cite{hs}. The first one, with metric
\beq\label{se1}
g_1=\frac{dx^2}{a^2}+\frac{d\phi^2}{a}\qq\qq a=\sqrt{x^4+c_2\,x^2+c_1\,x+c_0}\eeq
corresponds to the special case $\,\alf=l=m=n=0\,$ in Theorem 2. The second one had for metric
\beq\label{se2}
g_2=(a-x^2+p)\left(\frac{dx^2}{a^2}+\frac{d\phi^2}{a}\right),\eeq
where $\,p$ is some constant. This metric corresponds to the special case $\alf=l=m=n=0$ and 
$d=-p$ in our Theorem 3. It is difficult to push the comparison further since in these two references the existence of the quartic integral is proved but its explicit form is not given.
\item Tsiganov in \cite{Ts} has given an integrable system with a quartic integral. It does not 
belong to our family since there is no $\pth\,\pf$ term in his quartic integral. This is 
forbidden for us since it would imply that the conformal factor $\be$ vanishes identically. 
\erm

Let us turn ourselves to the study of the global structure and to the determination of the possible manifolds. In \cite{hs} the analysis was mostly interested in $M={\mb S}^2$, however the 
non-compact manifolds ${\mb H}^2$ and ${\mb R}^2$ do appear. 

The analysis (see \cite{Va}) is as follows: the positivity of $F$ requires for 
$x$ to be in some interval $(a,b)$. We will intensively use the scalar curvature 
to establish whether the boundaries $x=a$ and $x=b$ are apparent coordinate 
singularities or true singularities forbidding a manifold. In the absence of true singularities 
the nature of the manifold is then determined by establishing a global conformal transformation 
between the actual metric and the canonical metrics, given in Appendix A, for 
$\,{\mb S}^2$ and $\ {\mb H}^2$. 

Let us begin with the global analysis for the first case where $\mu$ vanishes.

\section{The global structure for $\mu=0$}
This section will cover the integrable systems of Theorem 2, for which the metric is
\beq\label{met1}
g=\be^2(x)\left(\frac{dx^2}{F}+\frac{d\phi^2}{\sqrt{F}}\right)\qq\qq\be^2(x)=b_0+\alf\,x\eeq
and we will write
\beq
F(x)=x^4+c_2\,x^2+c_0=(x^2+a)(x^2+\wti{a})\qq\qq G=\sqrt{F}-x^2-\frac 12(a+\wti{a}).
\eeq

\subsection{First case: $\alf=0$ and $c_1=0$}
Since $\,\be^2$ is constant we can set $b_0=1$. Let us first observe that the points 
$x=\pm \nf$ are apparent singularities of the metric (\ref{met1}). For instance $x=+\nf$ is 
mapped, by $u=1/x$, to $u=0+$ giving
\[g\sim du^2+u^2\,d\phi^2\]
which is an apparent coordinate singularity due to the use of polar coordinates. 

\nin Let us begin the global analysis with :

\begin{nth}
The integrable system in Theorem 2:
\brm
\item[(i)] Is trivial for $\wti{a}=a\, \in{\mb R}$.
\item[(ii)] Is not defined on a manifold if $\ \min(a,\wti{a})<0$ and $a\neq\wti{a}$. 
\item[(iii)] Is defined on ${\mb H}^2$ if $\wti{a}=0$ and $a>0$. It can be written \footnote{In all that follows we will use the shorthand notation $s=\sinh\chi,\ \  c=\cosh\chi\ $ for hyperbolic functions.}
\beq\label{Fcase1}\left\{\barr{l}\dst 
2\,H=P_v^2+\frac{c}{s^2}\,\pf^2+\ka\,\frac{s}{(c+1)^2}\,\cos\phi
-l\,\frac{s^2}{(c+1)^2}\\[4mm]\dst 
Q=\pf^4-4\ka\Big(\sin\phi\,P_v+\cos\phi\,\frac{\pf}{s}\Big)\pf+4l\,\pf^2
+\ka^2\,\frac{s^2}{(c+1)^2}\,\sin^2\phi.\earr\right.\eeq
\item[(iv)] Is defined on ${\mb S}^2$ if $\ 0<\wti{a}<a$. It can be written \footnote{In all that follows we will use the shorthand notation 
$S={\rm sn}\,(\tht,k^2),\,\ C={\rm cn}\,(\tht,k^2),\,\ D={\rm dn}\,(\tht,k^2)$ 
for Jacobi's elliptic functions.}
\beq\label{Fcase12}\left\{\barr{l}\dst 
2\,H=P_v^2+\frac{D}{S^2}\,\pf^2+\ka\,k^4\,\frac{SC}{(D+1)^2}\,\cos\phi
-l\,k^4\,\frac{S^2}{(D+1)^2}\qq k^2\in\,(0,1)\\[4mm]\dst 
Q=\pf^4-4\ka\Big(\sin\phi\,P_v+\frac CS\,\cos\phi\,\pf\Big)\pf+4l\,\pf^2
+\ka^2\,k^4\,\frac{S^2}{(D+1)^2}\,\sin^2\phi.\earr\right.
\eeq
\item[(v)] Is defined on ${\mb S}^2$ if $\ a\in\,{\mb C}\bs\{0\}$ and $\wti{a}=\ol{a}$. 
It can be written
\beq\label{Fcase13}\left\{\barr{l}\dst 
2\,H=P_v^2+\frac{\mu}{S^2}\,\pf^2-\ka\,k^2\,k'^2\,\frac{SC}{D^3}\,\cos\phi+
l\,k^2\,k'^2\,\frac{S^2}{D^2}\qq k^2\in\,(0,1)\\[4mm]\dst  
Q=\pf^4-\ka\left(\sin\phi\,P_v+\frac{C}{SD}\,\cos\phi\,\pf\right)\pf+l\,\pf^2-
\ka^2\,k^2\,k'^2\,\frac{S^2}{4\,D^2}\,\sin^2\phi
\earr\right.\eeq
with
\[\mu=1-k^2\,\frac{S^2\,C^2}{D^2}\geq k'^2.\]
\erm
\end{nth}

\nin{\bf Proof of (i):}

\nin In this case we have $\,G\equiv 0$ which implies, as observed in Section 2.1, that $\pf$ 
is conserved and so that $Q_4$ is reducible. Using the coordinate $u=1/x$ the metric
\[g=\frac{du^2}{(1+a\,u^2)^2}+\frac{u^2\,d\phi^2}{1+a\,u^2}\qq\quad a\in{\mb R},\]
is of constant scalar curvature since $R=2a$. The discussion is then 
\brm
\item For $a>0$ the change $\sqrt{a}\,u=\tan\tht$ gives the canonical metric (\ref{canS}) 
on ${\mb S}^2$. 
\item For $a=0$ we get ${\mb R}^2$ with its flat metric.
\item For $a<0$ the change  $\sqrt{|a|}\,u=\tanh\chi$ gives the canonical metric (\ref{canH1}) 
on ${\mb H}^2$. 
\erm
One can check that $Q$ itself is always fully reducible, trivializing the integrable system.
$\quad\Box$

\vspace{2mm}
\nin{\bf Proof of (ii):}

\nin Taking $a<\wti{a}$ we have to consider two cases:
\[(1):\quad a=-x_2^2,\quad \wti{a}=-x_1^2 \qq\qq (2):\quad a=-x_2^2,\quad\wti{a}=x_1^2
\quad\mbox{with}\quad 0<x_1<x_2.\]
In both cases positivity allows $x\in(x_2,+\nf)$ and we need to study the nature of the singularity at $x=x_2$. It can be ascertained from the scalar curvature
\[R=-\frac 34\,x_2(x_2^2+\wti{a})\,\frac 1{x-x_2}+O(1)\]
which shows that it is a true singularity of the metric, forbidding any manifold.

In the first case positivity also allows $x\in(-x_1,+x_1)$. In this case we have
\[R=-\frac 34\,x_2(x_2^2-x_1^2)\,\frac 1{x-x_1}+O(1)\]
leading to the same conclusion.$\quad\Box$

\vspace{2mm}
\nin{\bf Proof of (iii):}

\nin Here we have $a\in{\mb R}\bs\{0\}$. Up to a scaling of the observables, we can set $|a|=1$. 
Using again the coordinate $u=1/x$ the metric becomes
\beq\label{metric1}
g=\frac{du^2}{1+\eps\,u^2}+\frac{u^2}{\sqrt{1+\eps\,u^2}}\,d\phi^2\qq\qq \eps={\rm sign}\,(a).
\eeq
For $\eps=-1$ we have $u\in(0,1)$. However for $u\to 1-$ the scalar curvature 
\[R=-\frac 3{4(1-u)}+O(1)\]
exhibits a true singularity precluding any manifold.

For $\eps=+1$ we have $u\in (0,+\nf)$. As already observed 
$u=0+$ is an apparent singularity. The change of variable $u=\sinh v$  
gives for metric  and scalar curvature
\beq\label{metiii}
g=dv^2+\frac{\sinh^2 v}{\cosh v}\,d\phi^2\qq\qq R=1-\frac 32\,\tanh^2 v 
\qq v\in (0,+\nf)\quad \phi\ \mbox{azimuthal}\eeq
showing that the manifold is $\,M={\mb H}^2$. 

In the hamiltonian we can set $n=0$ but we must take $m=0$ for the potential 
to be defined on M. Transforming $\,H$ and $\,Q$ into the coordinates 
$\,(v,\,\phi,\,P_v,\,\pf)\,$ gives (\ref{Fcase1}).

Let us prove that this integrable system is globally defined on $M={\mb H}^2$. Writing 
the metric (\ref{metiii}) as
\[g=\frac{\sinh^2 v}{\cosh v}\left(d\phi^2+\frac{\cosh v}{\sinh^2 v}\,dv^2\right)\]
if we define a new coordinate $\chi$ by
\beq\label{nciii}
\frac{d\chi}{\sinh\chi}=\frac{\sqrt{\cosh v}}{\sinh v}\,dv\qq\qq 
v\in\,(0,+\nf)\ \to\ \chi\in\,(0,+\nf)
\eeq
we get, using the formulas of Appendix A:
\beq
g=\Om^2\,g(H^2,{\rm can})\qq\qq \Om^2=\frac{(1-t^2)^2\,\sinh^2 v}{4t^2\,\cosh v}\qq 
t\equiv\tanh\frac{\chi}{2}\ \in\,(0,1)
\eeq
where 
\beq\label{coord1}
t=\tanh \frac v2\,\,e^{\eta(v)}\qq\qq \eta(v)=-\int_v^{+\nf}\frac{(\sqrt{\cosh x}-1)}{\sinh x}\,dx.
\eeq
It follows that the first relation in (\ref{coord1}) can be extended to $v\in{\mb R}$. Then the 
function $t(v)$ is  odd (while $\eta(v)$ is even), $C^{\nf}$ and strictly increasing. 
This implies that its reciprocal function $v(t)$ is continuous, odd and strictly increasing 
for $t\in(-1,1)$. Since 
\[D_v\,t=\frac{\sqrt{\cosh v}}{1+\cosh v}\,e^{\eta(v)}\] 
never vanishes $v(t)$ is also $C^{\nf}$ for $t\in(-1,+1)$. 
It follows that $\eta\circ v(t)$ is an even continuous function of $t$ so we can define 
$\eta\circ v(t)=\psi(t^2)$. The function $\psi$ is $C^{\nf}$  for  $t^2\in(0,1)$ but we need 
to extend it to $[0,1)$. Since for $t\to 0+$ we have $v(t)\to 0+$, an easy expansion in powers 
of $v(t)$ shows that
\[\psi'\equiv D_{t^2}\psi=\frac{\eta'(v(t))}{D_v\,(t^2)}=c_0^{(1)}+c_1^{(1)}\,v(t)^2+O(v(t)^4).\]
This structure may be shown to hold for all the derivatives of $\psi$ with respect to 
the variable $\,t^2$ by recurrence. As a consequence we have  
\beq\label{tv1}
\tanh\frac v2=t\,e^{-\psi(t^2)}\qq\qq t\in[0,1)\eeq
and any $C^{\nf}$ function $f(v^2)$ for $v\in[0,+\nf)$ can be written $\wti{f}(t^2)$ 
which  will be a $C^{\nf}$ function of $t^2\in[0,1)$. 
Recalling that
\[t^2=\frac{\eta_3-1}{\eta_3+1}\]
the function $f(v^2)$ belongs to $C^{\nf}(M)$. Close to $t=0$ the Taylor expansion 
\beq\label{DLiii}
v=\tau-\frac{\tau^3}{24}+O(\tau^5)\qq\qq\tau=2e^{-\eta(0)}\,t
\eeq
will be useful.

We are now in position to prove that the integrable system given by (\ref{Fcase1}) is 
globally defined. Using the generators $M_i\ i=1,2,3$ in $\,T^*_M$ of the isometries 
one can write the hamiltonian
\beq
2H=\frac 1{\Om^2}\Big(M_1^2+M_2^2-M_3^2\Big)+ \ka\,\Lambda_1\,\eta_1-l\,\Lambda_2.\eeq
From the previous argument all the functions in $H$ belong to $C^{\nf}(M)$ as can be seen 
from their formulas   
\[\Lambda_1(t^2)=\frac{(1-t^2)\,\sinh( v(t))}{2t\,(\cosh(v(t))+1)^2}\qq\qq 
\Lambda_2(t^2)=\frac{\sinh^2(v(t))}{(\cosh(v(t))+1)^2}\]
and for $t\to 0+$ the check just uses relation (\ref{DLiii}).   
The cubic observable can be written
\beq
Q=M_3^4-4\ka\,\Big(\Lambda_3(t^2)\,M_1-\Lambda_4(t^2)\,\eta_1\,M_3\Big)M_3+\ka^2\,\Lambda_5(t^2)\,(\eta_2)^2+4l\,M_3^2.
\eeq
and by the same argument all the functions $\La_i(t^2)$ do belong to $C^{\nf}(M)$.$\quad\Box$

\vspace{2mm}
\nin{\bf Proof of (iv):}

\nin The function $\,F=(x^2+a)(x^2+\wti{a})\,$ is strictly positive and the metric is
\beq\label{met4}
g=\frac{dx^2}{F}+\frac{d\phi^2}{\sqrt{F}}\qq\qq x\in{\mb R}\qq \phi\ \ \mbox{azimuthal}
\eeq
with a scalar curvature
\[R=a+\wti{a}-\frac 32\,\frac{\de\,x^2}{F}\qq\qq \de=(a-\wti{a})^2\]
which is $C^{\nf}$ for all $x\in{\mb R}$. As already explained at the beginning of this section 
the points $x=\pm\nf$ are apparent singularities, geometrically the poles 
of a sphere. Since the $x$ coordinate gives untransparent expressions for $F,\,G$ and $\,G'$ we 
will use Jacobi elliptic functions. Let us consider the metric written in the variable 
$u=x^2\in(0,+\nf)$. We get first
\[g=\frac{du^2}{4u(u+a)(u+\wti{a})}+\frac{d\phi^2}{\sqrt{(u+a)(u+\wti{a})}}.\]
The change of variables 
\[u=a\,\frac{{\rm cn}\,^2(v,k^2)}{{\rm sn}\,^2(v,k^2)}\qq\qq \wti{a}=(1-k^2)\,a\qq k^2\in(0,1)\]
gives for the metric (no loss by taking $a=1$):
\beq
g=dv^2+\frac{S^2}{D}\,d\phi^2\qq\qq v\in(0,K)\qq \phi\ \ \mbox{azimuthal}.\eeq
using our earlier shorthand notations for Jacobi elliptic functions.  
However when going to the variable $u$ we have lost half of the manifold and this is why 
$\,v\in(0,K)$. But now we can recover the full manifold by extending $\,v\in(0,2K)$ since 
$v=0$ and $v=2K$ are indeed apparent singularities (the ``poles" of the manifold).  
The scalar curvature  
\[2R=3\,D^2-1-k'^2+3\,\frac{k'^2}{D^2}\]
is indeed $C^{\nf}([0,2K])$. In the hamiltonian we can set $n=0$ and we must impose 
$m=0$ since this piece is singular for $u\to 0+$. Then transforming $\,(H,Q)\,$ into the 
coordinates $(v,\,\phi,\,P_v,\,\pf)\,$ gives the formulas (\ref{Fcase12}). 

This integrable system is globally defined on $M={\mb S}^2$; to prove this let us define
\beq\label{tsphere}
t\equiv\tan\frac{\tht}{2}=\frac{S(v)}{C(v)+D(v)}\,e^{\eta(v)}\qq\qq  
\eta(v)=\int_K^v\,\frac{\sqrt{D(x)}-1}{S(x)}\,dx\eeq
which maps $v\in(0,2K)\to t \in (0,+\nf)$. As already explained in the proof of (iii) we have
\beq
g=\Om^2\ g(S^2,{\rm can})\qq\qq\Om^2(t^2)=\frac{(1+t^2)^2\,S^2(v(t))}{4t^2\,D(v(t))}.\eeq
The first relation in (\ref{tsphere}) can be extended to $v\in(-2K,2K)$ or $t\in{\mb R}$. 
Then $t(v)$ is odd, $C^{\nf}$ and strictly increasing. Hence $v(t)$ is odd, continuous 
and strictly increasing. Since 
\[t'=D_v\,t=\frac{1+\sqrt{D(v)}}{C(v)+D(v)}\,e^{\eta(v)}\] 
never vanishes $v(t)$ is in fact $C^{\nf}$ and  $\eta\circ v(t)$ is an even continuous function 
of $t$ so we can define $\eta\circ v(t)=\psi(t^2)$. By the same argument detailed in the 
proof of (iii) the function $\psi$ is $C^{\nf}$  for  $t^2\in[0,+\nf)$ and so any $C^{\nf}$ 
function $f(v^2)$ for $v\in[0,2K)$ can be written $\wti{f}(t^2)$. Since we have
\[t^2=\frac{1-\zeta_3}{1+\zeta_3}\] 
it follows that $\wti{f}(t^2)$ will be a $C^{\nf}$ function on the manifold, except at the south pole. This is not a problem since the relation
\beq\label{poleNS}
t(2K-v)=\frac 1{t(v)}\qq\Longrightarrow\qq v(1/t)=2K-v(t)\eeq
relates the behaviour at the north pole $(v=0;\,t=0;\,\,\zeta_3=+1)$ with the behaviour at 
the south pole $(v=2K;\,t=+\nf;\,\zeta_3=-1)$. So it will be sufficient to consider $t\in[0,+\nf)$ and analyze what happens at the north pole using the Taylor expansion
\beq\label{DLiv}
v(t)=\tau-\frac{(1+k'^2)}{24}\,\tau^3+O(\tau^5)\qq\qq \tau=e^{-\eta(0)}\,t.
\eeq 
Let us first examine the hamiltonian
\beq
2H=\frac 1{\Om^2(t^2)}\Big(L_1^2+L_2^2+L_3^2\Big)+\ka\,k^4\,\Lambda_1(t^2)\,\zeta_1-l\,k^4\,\Lambda_2(t^2)
\eeq
with
\[\Lambda_1(t^2)=\frac{(1+t^2)\,S(v(t))\,C(v(t))}{2t\,(D(v(t))+1)^2}\qq\qq\Lambda_2(t^2)=
\frac{S^2(v(t))}{(D(v(t))+1)^2}.\]
These three functions are $C^{\nf}$ for $t\in(0,+\nf)$ and upon use of (\ref{DLiv}) for $t\in[0,+\nf)$ 
they will belong to $C^{\nf}(M)$. For the quartic integral
\beq
Q=L_3^4-\ka\,\Big(\Lambda_3(t^2)\,L_1+\Lambda_4(t^2)\,\zeta_1\,L_3\Big)L_3
+\ka^2\,k^4\,\Lambda_5(t^2)\,(\zeta_2)^2.\qq 
\eeq
the argument is similar.$\quad\Box$

\vspace{2mm}
\nin{\bf Proof of (v):}

\nin The first change of coordinate $u=x^{-2}$ gives for metric
\beq
g=\frac{du^2}{4u\,p(u)}+\frac u{\sqrt{p(u)}}\,d\phi^2\qq\qq p(u)=(1+a\,u)(1+\ol{a}\,u)\qq u\in(0,+\nf)
\eeq
while the second change
\beq
{\rm cn}\,(2v)=\frac{1-|a|\,u}{1+|a|\,u}\qq\qq k^2=\frac 12\left(1-\frac{a+\ol{a}}{2\,|a|}\right)\qq
v\in(0,K)\eeq
transforms it into 
\beq
g=\frac 1{|a|}\left(dv^2+\frac{S^2}{\mu}d\phi^2\right)\qq\qq
\mu=1-k^2\frac{S^2\,C^2}{D^2}\geq \frac{2k'}{1+k'}.
\eeq
By a global scaling we can set $|a|=1$. The transition from $\,x\to u$ again loses half of the manifold. 
To recover it we will extend $\,v\in(0,2K)$ and $\phi$ azimuthal since both end-points $v=0$ and $v=2K$ are just the coordinate singularities corresponding to the ``poles"' of ${\mb S}^2$. The scalar curvature 
\[\frac R2=k'^2-k^2+\frac{12k'^4}{\mu^2}\,\frac{(2D^2-1)}{D^4}
-\frac{12k'^2}{\mu}\frac{D^2-k^2}{D^2}\]
is again $C^{\nf}([0,2K])$.

The computations of $G$ and 
$\sqrt{F}\,G'$ are tricky but give eventually the simple results
\[\sqrt{F}\,G'=-4\,k^2\,k'^2\,\frac{S\,C}{D^3}\qq\qq  G=2\,k^2\,k'^2\,\frac{S^2}{D^2}.\]
Again we have to set $m=n=0$ and the coordinates change   
$\,(x,\,\phi,\,P_x,\,\pf)\,\to\,(v,\,\phi,\,P_v,\,\pf)\,$ 
gives for $H$ and $Q$ the formulas (\ref{Fcase13}) up to a scaling of $\ka$.

To prove the global definiteness on $M={\mb S}^2$ let us give the key formulas needed. Defining
\beq
t\equiv\tan\frac{\tht}{2}=\frac{k'\,S(v)}{C(v)+D(v)}\,e^{\eta(v)}\qq
\eta(v)=\int_K^v\,\frac{\sqrt{\si(x)}-1}{S(x)}\,dx\qq 
\eeq
we have
\beq
g=\Om^2\,g(S^2,{\rm can})\qq\qq \Om^2(t^2)=\frac{(1+t^2)^2\,S^2(v(t))}{4t^2\,\si(v(t))}.\eeq
The relations (\ref{poleNS}) are still valid and the check of $C^{\nf}$-ness at the north pole needs
\[v(t)=\tau+\frac{(5k^2-1)}{12}\,\tau^3+O(\tau^5)\qq\qq \tau=e^{-\eta(0)}\,t.\]
Writing the hamiltonian as
\beq
2H=\frac 1{\Om^2(t^2)}\,\Big(L_1^2+L_2^2+L_3^2\Big)-\ka\,k^2\,k'^2\,\La_1(t^2)\,\zeta_1
+\ka\,k^2\,k'^2\,\La_2(t^2)\eeq
and the quartic integral 
\beq
Q=L_3^4-\kappa\Big(\La_3(t^2)\,L_1+\La_4(t^2)\,\zeta_1\,L_3\Big)L_3
-\ka\,k^2\,k'^2\,\La_5(t^2)\,(\zeta_2)^2\eeq
%\[\La_3=\frac{2t\,\sqrt{\si(v)}}{S(v)}\qq\La_4=\frac{(1+t^2)C(v)/D(v)
%-(1-t^2)\sqrt{\si(v)}}{2t\,S(v)}\qq\La_5=\frac{(1+t^2)^2\,S^2(v)}{4t^2\,D^2(v)}.\]
one can check that all the $\La_i(t^2)$ are $C^{\nf}$ functions for $t^2\in\,[0,+\nf[$ implying that the integrable system is globally defined on ${\mb S}^2$.
$\quad\Box$

\vspace{2mm}
\nin {\bf Remarks:}
\brm
\item The appearance of a constant curvature metric in case (i) was already observed in \cite{hs}. 
\item In \cite{hs} the case $F=x^4+a\,x^2+1$ is discussed in their Theorem 1. However their  
restriction  $\,a>-2$ for the hamiltonian to be globally defined on $M={\mb S}^2$ should be 
modified to $\,-2<a<2\,$.
\item If in the integrable system (\ref{Fcase12}) we take the limit $k^2\,\to\, 1$  we 
smoothly recover the system (\ref{Fcase1}). However the manifold changes drastically.
\item The cases (iv) and (v) can be fully analyzed in one stroke if one keeps the coordinate 
$\,x\,$. However the structure of the observables appears to be simpler using elliptic functions. 
\erm

\subsection{Second case: $\alf\neq 0$ and $c_1=0$}
Setting $\alf=1$ let us write the conformal factor $\be^2(x)=x-x_0$. The metric is
\beq\label{met1conf}
g=(x-x_0)\left(\frac{dx^2}{F}+\frac{d\phi^2}{\sqrt{F}}\right)\qq\qq F(x)=(x^2+a)(x^2+\wti{a}).\eeq
Let us first examine what is changing for  $x\to +\nf$. 
The metric becomes
\[g\sim\frac{dx^2}{x^3}+\frac{d\phi^2}{x}=4(d\tau^2+\tau^2\,d\Phi^2)\qq \tau=\frac 1{\sqrt{x}}\to 0+\qq 
\Phi=\frac{\phi}{2}\]
and if we take $\Phi$ to be azimuthal we see that $x=+\nf$ is again an apparent singularity. This gives $\,\cos(2\Phi)$ for the angular dependence of the potential. 

\begin{nth}
The integrable system in Theorem 2:
\brm
\item[(i)] Is trivial for $\wti{a}=a\,$.
\item[(ii)] Is not defined on a manifold if $\min(a,\wti{a})<0$ and $a\neq\wti{a}$.
\item[(iii)] Is defined on ${\mb H}^2$ if $\wti{a}=0$ and $a>0$. The hamiltonian becomes
\beq\label{Hcase2} 
2H=\frac 1{B}\left(A^2\,P_v^2+A\,\frac{\pF^2}{v^2}+\ka\,\frac{v^4}{(A+1)^2}\,\cos(2\Phi)
-l\,\frac{v^6}{(A+1)^2}+m+n\,v^2\right)\eeq
with 
\[A=\sqrt{1+v^4}\qq\qq B=1+\rho\,v^2\] 
and the quartic observable 
\beq\label{Qcase2}\barr{l}\dst 
Q=\pf^4-\frac{4\ka}{B}\,{\cal M}^2
+\frac{4\ka}{B}\,(B^2-2B+A)\,\cos^2\Phi\,\frac{\pF^2}{v^2}
+\frac{4\ka}{B}\,(B-A)\,\frac{\pF^2}{v^2}+4l\,\pF^2\\[4mm]\dst 
\hspace{8mm}-\frac{\ka^2}{B}\frac{(2-B)}{(A+1)^2}\,v^4\,\sin^2(2\Phi)
+4\frac{\ka}{B}\Big(\frac{v^2}{(A+1)^2}(\ka+l\,v^2)+\rho\,m-n\Big)\,v^2\,\sin^2\Phi\earr\eeq
with
\[{\cal M}=A\,\sin\Phi\,P_v+B\,\cos\Phi\,\frac{\pF}{v}.\]
\item[(iv)] Is not defined on a manifold for $\ 0<a<\wti{a}$ or $\wti{a}=\ol{a}$. 
\erm
\end{nth}

\nin{\bf Proof of (i):}

\nin Since $\dst G= \sqrt{F}-x^2-(a+\wti{a})/2$ vanishes identically $\pf$ is conserved so that 
$Q_4$ becomes reducible, trivializing the system.

\nin{\bf Proof of (ii):}

\nin Taking $a<\wti{a}$ we have to consider two cases:
\[(1):\quad a=-x_2^2,\quad \wti{a}=-x_1^2 \qq\qq (2):\quad a=-x_2^2,\quad\wti{a}=x_1^2
\quad\mbox{with}\quad 0<x_1<x_2.\]
In both cases positivity is ensured if we take $\,x\in(x_2,+\nf)$.
Let us consider again the metric behaviour at the end-point $x=x_2+$. If the conformal 
factor does not vanish for $x\to x_2+$ then it does not change 
the nature of the singularity, which remains a true one as already proved in Theorem 4 (ii). 
If the conformal factor vanishes as $x-x_2$  then the scalar curvature 
\[R=-\frac{3x_2}{4}(x_2^2+\wti{a})\,\frac 1{(x-x_2)^2}+O\left(\frac 1{x-x_2}\right)\]
shows that $x_2$ remains a true singularity.

In the first case positivity may be also satisfied with $\,x\in(-x_1,+x_1)$. If the conformal 
factor does not vanish for $x\to x_1-$ then it does not change 
the nature of the singularity, which remains a true one. If the 
conformal factor vanishes as $x-x_1$  then the scalar curvature 
\[R=-\frac{3x_2}{4}(x_2^2-x_1^2)\,\frac 1{(x-x_1)^2}+O\left(\frac 1{x-x_1}\right)\]
shows that $x_1$ remains singular.$\quad\Box$

\nin{\bf Proof of (iii):}

\nin In the metric (\ref{met1conf}) we can set $|a|=1$. The change of variable $x=1/v^2$, up 
to a global scaling, brings it into the form 
\beq
g=(1-x_0\,v^2)\left(\frac{dv^2}{1+\eps\,v^4}+\frac{v^2}{\sqrt{1+\eps\,v^4}}\,d\Phi^2\right)\qq 
\eps={\rm sign}\,(a)\qq\Phi=\frac{\phi}{2}\eeq
Let us first consider the case $\eps=-1$ which requires $v\in(0,1)$. The curvature is singular for $v\to 1-$ since we have
\[R=-\frac 3{2(1-x_0)}\,\frac 1{1-v}+O(1)\]
and for $x_0=1$ the singularity is even worse 
\[R=\frac 34\,\frac 1{(1-v)^2}+O\left(\frac 1{1-v}\right)\]
so there can be no manifold.

Let us consider now the case $\eps=+1$ for which  $v\in(0,+\nf)$ and $\Phi$ is an azimuthal angle. 
If $x_0>0$ let us define $x_0=1/\nu^2$; then positivity of the conformal factor requires 
$v\in\,(0,\nu)$ and the curvature is singular for $v \to \nu-$ since
\[R=-\frac{\nu(\nu^4+1)}{8}\,\frac 1{(v-\nu)^3}+O\left(\frac 1{(v-\nu)^2}\right)\]  
For $x_0\leq 0$ let us define $x_0=-\rho$. Now the conformal factor never vanishes and the 
metric becomes
\beq
g=(1+\rho\,v^2)\left(\frac{dv^2}{1+v^4}+\frac{v^2}{\sqrt{1+v^4}}d\Phi^2\right)\qq\qq
v\in(0,+\nf)\qq \Phi\quad\mbox{azimuthal}
\eeq
and $v=0$ is an apparent singularity. The scalar curvature 
\[R=-\frac{(\rho\,v^6-1)(\rho\,v^4+3v^2-2\rho)}{2(v^4+1)(1+\rho\,v^2)^2}\]
is now $C^{\nf}([0,+\nf))$.  

%The initial hamiltonian was given in Theorem 2 and the quartic observable is explicitly
%\beq\barr{l}
%Q=\pf^4+2\ka\Big(\alf\frac{F}{\be^2}\,\cos\phi\,P_x^2+2\,\sqrt{F}\,\sin\phi\,P_x\,\pf
%+(\alf\frac{\sqrt{F}}{\be^2}-2x)\,\cos\phi\,\pf^2\Big)+2\ka\,l\,\pf^2\\[4mm] 
%\hspace{1cm}+2\ka^2\left(\alf\,\frac{\sqrt{F}G'}{\be^2}\,\cos^2\phi-G\,\sin^2\phi\right)
%+\frac{2\ka}{\be^2}\Big(\alf\ka\,l\,G+\alf\,n-b_0\,m\Big)\cos\phi.\earr\eeq
Using for coordinates $(v,\,P_v,\,\Phi,\pF)$ in $T^*_M$ a lengthy computation leads to the 
integrable system given by (\ref{Hcase2}) and (\ref{Qcase2}), up to scalings and the elimination 
of a reducible piece proportional to the hamiltonian.

Let us now prove that the manifold is $M={\mb H}^2$ and that the system is globally defined on $M$. 
As already explained in the proof of Theorem 4 case (iii), if we define a new coordinate $t$ by
\beq\label{chi2}
t\equiv\tanh\frac{\chi}{2}=\frac v{1+\sqrt{1+v^2}}\,e^{\eta(v)}\qq 
\eta(v)=-\int_v^{+\nf}\left(\frac 1{\sqrt[4]{1+x^4}}-\frac 1{\sqrt{1+x^2}}\right)\frac{dx}{x}\eeq
mapping $v\in\,(0,+\nf)$ into $t\in\,[0,1)$ we have
\beq
g=\Om^2\,g(H^2,{\rm can})\qq\qq\Om^2=\frac{v^2}{4t^2}\frac{1+\rho\,v^2}{\sqrt{1+v^4}}(1-t^2)^2.
\eeq
The proof that $\eta\circ v(t)=\psi(t^2)$ is $C^{\nf}$ for $t\in[0,1)$ is fully similar to the one given in some detail for the proof of Theorem 4 case (iii). Writing the hamiltonian
\beq
2H=\frac 1{\Om^2}\Big(M_1^2+M_2^2-M_3^2\Big)+2\ka\,\La_1\,(\eta_1)^2-(\ka+l\,v^2)\,C+m+n\,v^2
\eeq
all functions of $v^2$ belong to $C^{\nf}(M)$ as well as
\[\La_1=C\,\frac{(1-t^2)^2}{4t^2}\qq\qq C=\frac{v^4}{(A+1)^2}\qq\qq A=\sqrt{1+v^4}.\]

The quartic first integral can be written
\[Q=M_3^4+\frac{4\ka}{B}\,{\cal M}^2-\frac{4\ka}{B}\,\La_2\,(\eta_1)^2\,M_3^2
+(\La_3+4l)M_3^2-\frac{4\ka^2}{b}\,\La_4\,(\eta_1\,\eta_2)^2+\frac{4\ka}{B}\,\La_5\,(\eta_2)^2\]
with
\[\La_2=\frac{(1-t^2)^2(B^2-2B+A)}{4v^2\,t^2}\qq \La_3=\frac{B-A}{v^2}\qq 
\La_5=\frac{v^2(1-t^2)^2}{4t^2}\left((k+l\,v^2)\frac C{v^2}+\rho\,m-n\right)\]
and 
\[\La_4=\frac{(1-t^2)^2(2-B)C}{4t^2}\qq 
{\cal M}=\frac{2t}{v}\frac{\sqrt{A}}{1-t^2}\left\{M_1+\frac{(1-t^2)^2}{4t^2}
\left(\frac{B}{\sqrt{A}}-\frac{1+t^2}{1-t^2}\right)\ \eta_1\,M_3\right\}\]
Taking into account that for $v\to 0+$ we have
\[B-A\sim \rho\,v^2\qq\qq B^2-2B+A\sim (\rho^2+1/2)v^4\qq\qq C\sim v^4/4\]
we conclude that ${\cal M}$ is globally defined and that all the functions $\La_i$ 
do belong to $C^{\nf}(M)$.$\quad\Box$

\nin{\bf Proof of (iv):}

\nin No matter how we reduce the interval for $x$ to ensure the positivity of the conformal factor it will begin or end at the zero of $x-x_0$. The vanishing of the  
conformal factor induces a true singularity as witnessed by the scalar curvature 
\[R=(x_0^2+a)(x_0^2+\wti{a})\,\frac 1{(x-x_0)^3}+O\left(\frac 1{(x-x_0)^2}\right)\]
so there is no manifold for this metric. $\quad\Box$

\subsection{Third case: $c_1\neq 0$}
In this case we have:

\begin{nth} For $c_1\neq 0$, no matter what $\alf$ is, the integrable models given by Theorem 2 
become globally trivial.
\end{nth}

\nin{\bf Proof:}

\nin The proofs of Theorems 4 and 5 have shown that $x=+\nf$ is always an apparent 
singularity which is either the ``pole" of $\,{\mb H}^2$ or one of the ``poles" of $\,{\mb S}^2$, when 
a manifold is available. Let us consider the second case, the analysis of the first case being fully similar. The coordinate $t$ on the sphere is given by
\[t\equiv\tan\frac{\tht}{2}=\exp\left(\eta_0-\int\frac{dx}{\sqrt[4]{F}}\right)\] 
and $x\to +\nf$ is mapped into $t \to 0+$. One gets
\[\tau\equiv e^{-\eta_0}\,t=\frac 1x+O\left(\frac 1{x^3}\right)\qq\Longrightarrow\qq 
\frac 1x=e^{-\eta_0}\,t+O(t^3)\]
Let us consider in $Q$ the term
\[\ka^2\,G\,\sin^2\phi=4\ka^2\,G\,\sin^2\Phi\,\cos^2\Phi=\ka^2\,\frac{(1+t^2)^2\,G}{t^2}\ (\ze_1\,\ze_3)^2.\]
The relations
\[G=\frac{c_1}{2x}+O\left(\frac 1{x^2}\right)\qq\Longrightarrow\qq 
\frac{G}{t^2}=\frac{c_1\,e^{-\eta_0}}{2t}+O(1)\]
show that this term is not even defined for $t=0$ so we must impose $\ka=0$. Then $\,\pf$ becomes conserved and $Q_4$ is trivialized.

Up to now we have just proved that any interval which includes $\pm\nf$ will not give a 
globally defined integrable system. However there remains the possibility that the manifold 
be recovered for some bounded interval $(x_2,x_1)$ where $F$ vanishes at both end-points.
 
So let us first consider the case where $x_1$ is a simple zero and $\alf=0$ hence $\be^2=1$. 
We can write $F=(x_1-x)P(x)$ where $P$ is a cubic polynomial such that $\,P(x_1)\neq 0$. The curvature 
\[R=-\frac 38\,\frac{P(x_1)}{x_1-x}+O(1)\]
is singular for $x\to x_1-$.

If $\alpha\neq 0$ either $\be^2=x_0-x$ does not vanish for $x=x_1$ and the curvature remains singular or $x_0=x_1$ and then the singularity is even worse
\[R=\frac 38\,\frac{P(x_1)}{(x_1-x)^2}+O\left(\frac 1{x_1-x}\right)\]
If $x_1$ is threefold there cannot be a bounded interval of positivity for $F$ and  we are left with the single possibility that the zeroes $x_1$ and $x_2$ be twofold with $F=(x-x_1)^2(x-x_2)^2$. Since the third power of $x$ is absent from $F$ we get $x_2=-x_1$ leading 
to $F=(x^2-x_1^2)^2$ for which $c_1=0$, a contradiction. $\quad\Box$

\section{The global structure: $\mu\neq 0$}
We will now consider the integrable systems of Theorem 3. Let us begin with the simpler case 
where $r=0$.

\subsection{First case: $r=0$}
The metric becomes
\beq
g=\frac{p(x)}{(x^2+d)^2}\left(\frac{dx^2}{x^2+d}+d\phi^2\right)\qq p(x)=\alf\,x^3+c_2\,x^2+c_1\,x+c_0\eeq
By a scaling we can set $\,d=\eps$ with either $\,\eps=0$ or $\,\eps=\pm 1$. Switching to 
$u=1/x$ we get
\beq
g=\frac{u^2\,p(1/u)}{1+\eps\,u^2}\left(\frac{du^2}{(1+\eps\,u^2)^2}
+\frac{u^2\,d\phi^2}{1+\eps\,u^2}\right).\eeq
It follows that we have:

\begin{nth}
The integrable system described by Theorem 3:
\brm
\item[(i)] For $\eps=0$ is defined on $M={\mb R}^2$ and becomes
\beq\label{dzero}\left\{\barr{l}\dst 
2H=\frac 1B\Big(P_x^2+P_y^2+2\,\ka\,x+l\,(x^2+y^2)+m\Big)\\[4mm]\dst 
Q=2\rho\,H\,L_3^2-2\ka\,P_y\,L_3-l\,L_3^2+\ka^2\,y^2\qq\qq L_3=x\,P_y-y\,P_x\earr\right.\eeq
with
\[B=1+\rho(x^2+y^2)\qq\qq\rho> 0\]
\item[(ii)] For $\eps=-1$ is defined on $M={\mb H}^2$ and becomes
\beq\label{dmoinsun}\left\{\barr{l}\dst 
2H=\frac 1B\Big(M_1^2+M_2^2-M_3^2+2\ka\,\eta_1\,\eta_3+l\,\eta_3^2+m\Big) \\[4mm]\dst 
Q=4\rho\,H\,M_3^2-2\ka\,M_1\,M_3-l\,M_3^2+\ka^2\,\eta_2^2\earr\right.\eeq
with
\[B=1-\rho+2\rho\,\eta_3^2\qq\qq\qq \rho \in\,(-1,0)\cup (0,+\nf).\]
\item[(iii)] For $\eps=+1$ is defined on $M={\mb S}^2$ and becomes
\beq\label{dplusun}\left\{\barr{l}\dst 
2H=\frac 1B\Big(L_1^2+L_2^2+L_3^2+2\ka\,\ze_1\,\ze_3+l\,\ze_3^2+m\Big)\\[4mm]\dst 
Q=4\rho\,H\,L_3^2+2\ka\,L_1\,L_3-l\,L_3^2-\ka^2\,\ze_2^2\earr\right.\eeq
with
\[B=1-\rho+2\rho\,\ze_3^2\qq\qq\qq \rho \in (-1,0)\cup (0,+1).\]
\erm

\end{nth}

\vspace{5mm}
\nin{\bf Proof of (i):}

\nin For $\eps=0$ we have
\[g=\left(\frac{\alf}{u}+c_2+c_1\,u+c_0\,u^2\right)\,g_0\qq\qq 
g_0=du^2+u^2\,d\phi^2\]
Taking $\,u\in(0,+\nf)$ and $\,\phi$ azimuthal $g_0$ is just the canonical flat metric 
on $\,{\mb R}^2$. Going back to cartesian coordinates we have $u=\sqrt{x^2+y^2}$ therefore 
we must take $\alf=c_1=0$. The potential $m\,x=m/u$ is not globally defined so we take 
$m=0$ and change $n\to m$. Then transforming back the observables to cartesian coordinates 
gives the desired result up to scalings.$\quad \Box$

\vspace{5mm}
\nin{\bf Proof of (ii):}

\nin For $\eps=-1$ the change of coordinates $u=\tanh\chi$, up to an argument similar 
to the case $\eps=+1$ which gives $\alf=c_1=0$, brings the metric into the form  
\[g=\la\Big(1+\rho\,\cosh(2\chi)\Big)\,g(H^2,{\rm can})
\qq\la=\frac{c_2-c_0}{2}\qq\rho=\frac{c_2+c_0}{c_2-c_0}\]
and we will write the conformal factor

where $\rho$ has to be appropriately constrained to prevent the vanishing of the conformal factor. 
Again we must have $m=0$ and change $n\to m$. The final form of the quartic integral is obtained 
up to scalings.$\quad \Box$

\vspace{5mm}
\nin{\bf Proof of (iii):}

\nin For $\eps=1$ the change of coordinates $u=\tan\tht$ brings the metric into the form
\[g=\Big(\alf\frac{\cos^3\tht}{\sin\tht}+\frac{c_2+c_0}{2}+\frac{c_1}{2}\,\sin(2\tht)+\frac{c_2-c_0}{2}\,\cos(2\tht)\Big)\,g(S^2,{\rm can})\qq\tht\in(0,\pi).\]
One must impose $\alf=0$ and since $\sin\tht$ is not differentiable at the poles $c_1=0$. 
The conformal factor becomes
\[\la\,\Big(1+\rho\,\cos(2\tht)\Big)\qq\qq\la=\frac{c_2+c_0}{2}\qq\rho=\frac{c_2-c_0}{c_2+c_0}\]
and the parameter $\rho$ is constrained in order to avoid a zero which would induce a 
curvature singularity. The final form of the observables, setting $m=0$, changing $n\to m$ and some scalings is then easily obtained.$\quad \Box$

In all of the following Sections we will set $r\neq 0$.

\subsection{Second case: $\alf=c_1=0$}
In this case $\,p=c_2\,x^2+c_0\,$ and changing $d\ \to\ \de$ we have
\beq
F=x^4+(2\de-4\,r\,c_2)x^2+\de^2-4\,r\,\,c_0\qq\qq \be^2=\frac 1{2\,r}(x^2+\de-\sqrt{F}).\eeq
Defining 
\[a+\wti{a}=2\de-4\,r\,c_2\qq\qq a\,\wti{a}=\de^2-4\,r\,c_0\]
we have
\[F=(x^2+a)(x^2+\wti{a})\qq\qq 2r\,G=\sqrt{F}-x^2-\frac{a+\wti{a}}{2}\]
and we can state:

\begin{nth}
The integrable system described by Theorem 3:
\brm
\item[(i)] Is trivial for $\wti{a}=a\, \in{\mb R}$.
\item[(ii)] Is not defined on a manifold if $\ \min(a,\wti{a})<0$ and $a\neq\wti{a}$. 
\item[(iii)] Is defined on ${\mb H}^2$ if $\wti{a}=0$ and $a>0$. It can be written
\beq\label{FFcase1}\left\{\barr{l}\dst 
2\,H=\frac{\eps}B\left(P_v^2+\frac{c}{s^2}\,\pf^2+\ka\,\frac{s}{(c+1)^2}\,\cos\phi
-\ka\,l\,\frac{s^2}{(c+1)^2}+m\right)\qq \eps=\pm 1\\[4mm]\dst 
Q=\pf^4+4\,\eps\,H\,\pf^2-4\ka\Big(\sin\phi\,P_v+\frac{\cos\phi}{s}\,\pf\Big)\pf+4\ka\,l\,\pf^2
+\ka^2\,\frac{s^2}{(c+1)^2}\,\sin^2\phi\earr\right.\eeq
with
\[B=\de-\frac 1{c+1}.\]
For $\eps=-1$ and $\de=0$ the manifold becomes ${\mb S}^2$ and the integrable 
system is Kovalevskaya's given by (\ref{KovaH}) and (\ref{KovaQ}).
\item[(iv)] Is defined on ${\mb S}^2$ if $\ 0<\wti{a}<a$. It can be written 
\beq\label{FFcase12}\left\{\barr{l}\dst 
2\,H=\frac{\eps}{B}\left(P_v^2+\frac{D}{S^2}\,\pf^2+\ka\,k^4\,\frac{SC}{(D+1)^2}\,\cos\phi
-\ka\,l\,k^4\,\frac{S^2}{(D+1)^2}+m\right)\quad \eps=\pm 1\quad k^2\in\,(0,1)\\[4mm]\dst 
Q=\pf^4+4\,\eps\,H\,\pf^2-4\ka\Big(\sin\phi\,P_v+\frac CS\,\cos\phi\,\pf\Big)\pf+4\ka\,l\,\pf^2
+\ka^2\,k^4\,\frac{S^2}{(D+1)^2}\,\sin^2\phi\earr\right.
\eeq
with
\[B=\de-1+\frac{k^2}{D+1}.\]
\item[(v)] Is defined on ${\mb S}^2$ if $\ a\in\,{\mb C}\bs\{0\}$ and $\wti{a}=\ol{a}$. 
It can be written
\beq\label{FFcase13}\left\{\barr{l}\dst 
2\,H=\frac{\eps}{B}\left(P_v^2+\frac{\si}{S^2}\,\pf^2-\ka\,k^2\,k'^2\,\frac{SC}{D^3}\,\cos\phi+
\ka\,k^2\,k'^2\,\frac{S^2}{D^2}+m\right)\quad \eps=\pm 1\quad k^2\in\,(0,1)\\[4mm]\dst  
Q=\pf^4+4\,\eps\,H\,\pf^2-\ka\left(\sin\phi\,P_v+\frac{C}{SD}\,\cos\phi\,\pf\right)\pf+\ka\,l\,\pf^2-
\ka^2\,k^2\,k'^2\,\frac{S^2}{4\,D^2}\,\sin^2\phi
\earr\right.\eeq
with
\[B=\de+1-2\frac{k'^2}{D^2}.\]
\erm
\end{nth}
We use again the shorthand notations for hyperbolic and elliptic functions given in Theorem 4.

\vspace{5mm}
\nin{\bf Proof of (i):}

\nin In this case $G\equiv 0$ hence $Q_4$ is reducible and the system is trivial. $\quad\Box$

\vspace{5mm}
\nin{\bf Proof of (ii):}

\nin Taking $a<\wti{a}$ we have to consider two cases:
\[(1):\quad a=-x_2^2\qq \wti{a}=-x_1^2\quad 0<x_1<x_2\qq\qq (2):\quad a=-x_2^2<0\qq\wti{a}=x_1^2.\]
In both cases positivity allows $x\in(x_2,+\nf)$. If $\be^2(x_2)\neq 0$, as already seen in 
the proof of Theorem 4, case (ii), $x=x_2$ remains a true singularity. If $\be^2(x_2)=0$ 
then $x=x_2$ remains a curvature singularity since
\[R=\frac{r\,x_2}{2(x-x_2)}+O(1).\]
In the first case we may also take $x\in\,(-x_1,+x_1)$. If $\be^2(x_1)\neq 0$, as already seen in 
the proof of Theorem 4, case (ii), $x=x_1$ remains a true singularity. If $\be^2(x_1)=0$ 
the curvature is  
\[R=\frac{r\,x_1}{2(x-x_1)}+O(1)\]   
and $x=x_1$ remains a curvature singularity $\quad\Box$

\vspace{5mm}
\nin{\bf Proof of (iii):} 

\nin Defining $x=1/u$ and setting $\eps=\pm 1$ for the sign of $r$, the metric is homothetic to
\[g=\eps\,B\left(\frac{du^2}{1+a\,u^2}+\frac{u^2\,d\phi^2}{\sqrt{1+a\,u^2}}\right)\qq\quad  
B=\de-\frac 1{1+\sqrt{1+a\,u^2}}\qq\qq a=\pm 1\]
For $a=-1$ the curvature
\[R=-\frac 34\,\frac 1{\de-1}\,\frac 1{1-u}+O\left(\frac 1{\sqrt{1-u}}\right)\]
is singular for $u\to 1-$ and it remains singular for $\de=1$ since in this case we have 
\[R=-\frac 14\,\frac 1{1-u}+O(1).\]
So we must consider $a=+1$ and the change of variable $u=\sinh v$ gives for the metric
\[g=\eps\,\left(\de-\frac 1{c+1}\right)\left(dv^2+\frac{s^2}{c}\,d\phi^2\right)
\qq v\in\,(0,+\nf)\qq \phi\ \mbox{azimuthal}\]
using the notations of Theorem 4, case (iii). 

If the conformal factor never vanishes for $v\geq 0$ we will get for manifold $M={\mb H}^2$. 
We have therefore two cases:
\beq
\Big(\eps=+1\qq \de>\frac 12\quad\Rightarrow\quad B>0\Big)\qq\mbox{or}\qq 
\Big(\eps=-1\qq\de<0\quad\Rightarrow\quad B<0\Big).
\eeq
In these two cases the global definiteness proof needs not be repeated since it follows closely Theorem 4, case (iii).

A very unusual manifold bifurcation appears for $\eps=-1$ and $\de=0$. The volume is 
{\em finite} with a $C^{\nf}([0,+\nf))$ curvature. To understand what is the true manifold, 
let us make the coordinate change $\dst \sinh(v/2)=\tan\tht$ 
mapping $\,v\in\,(0,+\nf)\ \to\ \tht\in\,(0,\pi/2)$. The metric becomes
\[g=2\left(d\tht^2+\frac{\sin^2\tht}{1+\sin^2\tht}\,d\phi^2\right)\qq\qq\qq  
R=2\frac{(2-\sin^2\tht)}{(1+\sin^2\tht)^2}\]
which would be a metric on $P^2({\mb R})$. However we can extend the range of $\,\tht$ to 
$\,(0,\pi)$ and the points $\tht=0$ and $\tht=\pi$ are just the apparent coordinate 
singularities giving for manifold $\,{\mb S}^2$. We recognize this 
metric as being Kovalevskaya's and a simple computation shows that $H$ and $Q$ are 
indeed given by (\ref{KovaH}) and (\ref{KovaQ}) since  one must set $l=m=0$ to avoid 
the singularity on the ``equator" $\,\tht=\pi/2$. $\quad\Box$

\vspace{5mm}
\nin{\bf Proof of (iv) and (v):} 

\nin Here too there is no need to repeat the proofs given in Theorem 4, cases (iv) and (v), 
provided that the conformal factors do not vanish for $v\in[0,2K]$. 
The resulting constraints for the case (iv) are 
\beq
\Big(\eps=+1\qq \de>1-\frac{k^2}{2}\quad\Rightarrow\quad B>0\Big)\qq\mbox{or}\qq 
\Big(\eps=-1\qq\de<k'\quad\Rightarrow\quad B<0\Big)
\eeq
and for the case (v) 
\beq
\Big(\eps=+1\qq \de>1\quad\Rightarrow\quad B>0\Big)\qq\mbox{or}\qq 
\Big(\eps=-1\qq\de<1-2k^2\quad\Rightarrow\quad B<0\Big).
\eeq
They are easily obtained since $B$ is monotonous.$\quad\Box$

\subsection{Third case: $\alf\neq 0$ }
We have for $F$ the structure   
\beq
F\equiv (x^2+d)^2-4r\,p=x^4+a_3\,x^3+a_2\,x^2+a_1\,x+a_0\qq\quad a_3=-4r\alf.\eeq
Since $a_3\neq 0$ no matter whether $\,a_1=-4r\,c_1$ vanishes or not $F$ is 
the most general quartic polynomial.

The relevant functions are
\beq
\be^2=\frac 1{2r}\Big(d+x^2-\sqrt{F}\Big)\qq G=\frac 1{4r}\Big(-a_2+\frac{a_3^2}{4}-a_3\,x-2\,x^2+2\,\sqrt{F}\Big)\eeq
and the metric
\beq\label{metfin}
g=\be^2\left(\frac{dx^2}{F}+\frac{d\phi^2}{\sqrt{F}}\right).\eeq

Let us begin with two preparatory lemmas:

\begin{nlem}
The points $x=\pm \nf$ are apparent singularities of the metric (\ref{metfin}). The metric 
may be defined on a manifold for $x\in\,(x_0,+\nf)$ with $F(x_0)=0$ and provided that $F$ 
be strictly positive on this interval. A simple zero or a fourfold zero of $F$ are excluded.
\end{nlem}

\vspace{5mm}
\nin{\bf Proof:} 

\nin For $x\to +\nf$ we have
\[g\sim \frac{-a_3}{r}\left(\frac{dx^2}{4x^3}+\frac{d\phi^2}{4x}\right)=
\frac{-a_3}{r}\Big(d\tau^2+\tau^2\,d\Phi^2\Big)\qq\quad \tau=\frac 1{\sqrt{x}}\to 0+  
\qq \Phi=\frac{\phi}{2}\]
and if we take $\Phi$ to be azimuthal this singularity is apparent. 

For $x\to -\nf$ we have 
\[g\sim \frac{-a_3}{r}\left(\frac{dx^2}{4x^3}+\frac{d\phi^2}{4x}\right)=
\frac{a_3}{r}\Big(d\tau^2+\tau^2\,d\Phi^2\Big)\qq\quad \tau=\frac 1{\sqrt{-x}}\to 0+\]
and we are led to the same conclusion. The change of sign of the metric from $x\to +\nf$ to 
$x\to -\nf$ forbids $x\in\,{\mb R}$, hence there must be a zero $x_0$ for which $F$ 
vanishes but remains strictly positive in $(x_0,+\nf)$.

Let it be supposed that the zero of $F$ is simple. We can write
\[F=u\,P(u) \qq u\equiv x-x_0 \qq\qq P(u)=u^3+b_2u^2+b_1u+b_0\qq b_0>0.\]
It follows that the product of the roots of $P$ is positive hence we will have at least 
one positive root, which is to be excluded. 

The case where the zero is fourfold is also excluded since then $G$ vanishes identically. $\quad\Box$

This lemma covers the case of an infinite range for $x$. However positivity allows also a 
finite range of $x$ between two zeroes of $F$. This is considered in

\begin{nlem}
If $F$ is strictly positive between two of its zeroes $(x_1,x_2)$ the metric 
(\ref{metfin}) may be defined on a manifold only if $x_1$ and $x_2$ are twofold. 
\end{nlem}

\vspace{5mm}
\nin{\bf Proof:} 

\nin Let $x=x_1$ and $x=x_2>x_1$ be simple zeroes of $F$. We have 
\[F(x)=(x-x_1)(x_2-x)\,P(x) \qq\qq\be^2=\frac 1{2r}\,B\qq B=d+x^2-\sqrt{F}\]
where $P(x)$ must be strictly positive for $x\in \, [x_1,x_2]$. The curvature, 
omitting the factor $1/2r$ in $\be^2$ is singular at both end-points:
\[R=\frac 38\,\frac{(x_1-x_2)}{(d+x_1^2)}\,\frac{P(x_1)}{x-x_1}+O(1)\qq 
R=-\frac 38\,\frac{(x_1-x_2)}{(d+x_2^2)}\,\frac{P(x_2)}{x-x_2}+O(1)\]
provided that $B$ does not vanish in $\,[x_1,\,x_2]$. If $B(x_1)=0$ the curvature remains singular
\[R=\frac{x_1}{4}\frac 1{x-x1}+O(1)\]
for $x=x_1$.

Alternatively we can take $x_1$ to be twofold $\,F=(x-x_1)^2(x_2-x)P(x)$ with $P>0$ for $x\in\,[x_1,x_2]$ in which case the curvature becomes continuous at $x=x_1$ but remains singular at $x=x_2$ for all possible  values of $d$. However for the special case $d=x_2=0$ the singularity for $x=x_2$ disappears leaving a continuous curvature. This special case is therefore given by
\[F=(x-x_1)^2\,(x^2-b\,x)\qq d=x_2=0\qq b>0\qq x\in\,(x_1,0)\]
with the metric
\[{\cal B}\left(\frac{dx^2}{(x-x_1)^2(b-x)\sqrt{-x}}+\frac{d\phi^2}{(x-x_1)\sqrt{b-x}}\right)
\qq {\cal B}=(-x)^{3/2}-(x-x_1)\sqrt{b-x}\]
but one has
\[{\cal B}(x_1)=(-x_1)^{3/2}>0\qq\mbox{and}\qq {\cal B}(0)=-\sqrt{b}\,(-x_1)<0\]
showing that ${\cal B}(x)$ vanishes in $(x_1,0)$ leading to a curvature singularity.

The last possible case is $\,F(x)=(x-x_1)^2(x_2-x)^2$ giving a continuous curvature at both end-points.
$\quad\Box$

These results will be used to prove:

\begin{nth}
The locally integrable system described in Theorem 3:
\brm
\item[(i)] Is globally defined on $M={\mb H}^2$ if $F$ has a twofold zero and $\de\neq 0$. 
Its hamiltonian is
\beq\label{firstH} 
2\,H=\frac 1B\left(F\,P_v^2+\sqrt{F}\,\frac{\pF^2}{v^2}+\ka\sqrt{F}\,v^2\,G_{,v^2}\,\cos(2\Phi)
+l\,v^2\,G+m+n\,v^2\right)
\eeq
where $v\in\,(0,+\nf)$ and $\Phi$ is azimuthal, with
\[F=1+2\si\,v^2+v^4 \qq \si\in\,(-1,+1)\qq G=\frac{\sqrt{F}-1-\si\,v^2}{v^4}-\frac{(1-\si^2)}{2}\]
and
\[B=\rho+\de\,v^2-\frac{\sqrt{F}-1-\si\,v^2}{v^2}\qq\qq  
{\cal M}=\sin\Phi\,\sqrt{F}\,P_v+\frac B{\rho}\,\cos\Phi\,\frac{\pF}{v}.\]
The quartic integral is
\beq\label{firstQ}\barr{l}\dst Q=\frac 14\,\pF^4+H\,\pF^2+\ka\frac{\rho}{B}\,{\cal M}^2
+\ka\left[\si+\frac 1{v^2}\left(2-\frac B{\rho}-\frac{\rho}{B}\sqrt{F}\right)\right]\cos^2\Phi\,\pF^2
\\[4mm]\dst  
-\ka\left[\frac{\si}{2}+\frac 1{v^2}\left(1-\frac{\rho}{B}\sqrt{F}\right)\right]\,\pF^2
+\frac l2\,\pF^2+\frac{\ka^2}{2}\left(\frac{\rho}{B}\,\sqrt{F}\,v^2\,G_{,v^2}-G\right)\,\sin^2(2\Phi)
\\[4mm]\dst 
+\ka\,\frac{\rho}{B}\left(-\ka\,\sqrt{F}\,v^2\,G_{,v^2}+l\,v^2\,G+m\frac{\rho-B}{\rho}+n\,v^2\right)\,\sin^2\Phi\earr
\eeq
The parameters $\rho$ and $\de$ must ensure that $B\neq 0$ for $v\geq 0$.

\item[(ii)] Is globally defined on $M={\mb S}^2$ if $F$ has a twofold zero and $\,\de=0$. 
Its hamiltonian is
\beq\label{Hdelta0}
2H=\frac 1B 
\left(P_v^2+\mu\,\frac{\pF^2}{S^2}+2\ka\,k^2\,k'^2\frac{(D^2-2k'^2)}{D^4}\,S^2\,\cos(2\Phi)+m\right)
\eeq
with
\[\mu=1-k^2\frac{S^2\,C^2}{D^2}\qq\qq B=\rho-2k^2\,k'^2\frac{S^2}{D^2}\]
and the quartic integral
\beq\label{Qdelta0}\barr{l}\dst 
Q=\frac 14\,\pF^4+H\,\pF^2+\ka\frac{\rho}{B}\,{\cal L}^2
-2\ka\frac{k^2\,k'^2}{B}\left(\rho+1-2k^2+2\frac{k^2\,k'^2}{\rho}\frac{C^2}{D^4}\right)\frac{S^2}{D^2}\cos^2\Phi\,\pF^2\\[4mm]\dst 
-\frac 1{B}\left((k^2-3/2)B+2\rho\,\frac{k'^2}{D^2}
+2k^2k'^2\frac{C^2-S^2\,D^2}{D^4}\right)\pF^2\\[4mm]\dst
-2\ka^2(k^2k'^2)^2\left(\frac{\rho}{2}-k^2+\frac{k'^2}{D^2}\right)\frac{S^4}{D^4}\,\sin^2(2\Phi)
+2\ka\frac{k^2k'^2}{B}\left(m-\rho\ka\frac{(D^2-2k'^2)}{D^2}\right)\frac{S^2}{D^2}\sin^2\Phi
\earr\eeq
with
\[{\cal L}=\sin\Phi\,P_v+\frac{B}{\rho}\,\frac{C}{SD}\cos\Phi\,\pF\]
The parameters $\rho$ and $\de$ must ensure that $B\neq 0$ for $v\in\,[0,2K]$.

\item[(iii)] Is globally defined in ${\mb H}^2$ if $F$ has a threefold zero and $\de\neq 0$. Its hamiltonian is
\beq\label{tripleH}
2H=\frac 1B\left(P_v^2+\frac c{s^2}\pF^2+\ka\,\frac{s^2}{(c+1)^3}\,\cos(2\Phi)+l\,\frac{(c+3)s^4}{(c+1)^3}+m\,s^2\right)
\eeq
and the quartic integral
\beq\label{tripleQ}\barr{l}\dst 
Q=\frac 14\pF^4+H\,\pF^2+\frac{2\rho\ka}{B}\,{\cal M}^2
+\frac{2\ka}{B}\left(\frac{\rho}{c+1}-\De-\frac B4\right)\pF^2+4\,l\,\pF^2\\[5mm]\dst 
+\frac{2\ka}{B}\left(\frac{\De}{c+1}-\frac{\De^2}{\rho}+\frac B{2(c+1)^2}\right)\,s^2\,\cos^2\Phi\,\pF^2 
-\frac{\ka^2}{B(c+1)^3}\left(\De+\frac B{4(c+1)}\right)s^4\,\sin^2(2\Phi)\\[5mm]\dst 
+2\frac{\rho\ka}{B}\left(-\frac{\ka}{(c+1)^3}+l\frac{(c+3)s^2}{(c+1)^3}+m\,\right)s^2\,\sin^2\Phi
\earr\eeq
with
\[B=\rho+\De\,s^2\qq\qq \De=\de+\frac 1{2(c+1)^2}\qq\qq {\cal M}=\sin\Phi\,P_v
+\frac B{\rho}\,\cos\Phi\,\frac{\pF}{s}.\]
The parameters $\rho$ and $\de$ must ensure that $B\neq 0$ for $v\geq 0$.

\item[(iv)] Is globally defined in ${\mb S}^2$ if $F$ has a threefold zero and $\de=0$ and 
$\rho=-1/2$. Its hamiltonian is
\beq\label{GoH}
2H=\pth^2+\frac{(1+\sin^2\tht)}{\sin^2\tht}\,\pF^2+\ka\,\sin^2\tht\,\cos(2\Phi)
\eeq
and its quartic integral
\beq\label{GoQ}\barr{l}
Q=\pF^4-2H\,\pF^2+\ka\,{\cal L}^2+\ka\Big(1+\cos^2\tht(1-\sin^2\tht\,\cos^2\Phi)\Big)\,\pF^2
\\[4mm] 
\hspace{8cm}+\ka^2\sin^2\tht\,\sin^2\Phi(\sin^2\tht\,\cos^2\Phi-1)\earr
\eeq
with
\[{\cal L}=\sin\Phi\,\pth+\cos^2\tht\,\frac{\cos\Phi}{\tan\tht}\,\pF\]
which is nothing but a special case of Goryachev system.
\erm
\end{nth}

\vspace{5mm}
\nin{\bf Proof of (i):} 

\nin $F$ has a twofold zero for $x=x_1$. Using $u=x-x_1$ we can write $F=u^2(u^2+b_1 u+b_0)$ with $b_0>0$ and $u\in\,(0,+\nf)$. We will use as a new variable $u=\sqrt{b_0}/v^2$ with $\,v\in\,(0,+\nf)$ 
which results in
\[\barr{lll}\dst \sqrt{F}=b_0\frac{\sqrt{\Psi}}{v^4}\qq & \dst \Psi(v^2)=1+2\si\,v^2+v^4\qq & \qq\dst 
\si=\frac{b_1}{2\sqrt{b_0}}\\[4mm]\dst \be^2=\frac{b_0}{2r}\,\frac B{v^2}\qq & \dst B(v^2)=\rho+\de\,v^2-\frac{\sqrt{\Psi}-1-\si\,v^2}{v^2} &\dst  \\[4mm]\dst   
G=\frac{b_0}{2r}\,\Ga & \dst\Ga=\frac{\sqrt{\Psi}-1-\si\,v^2}{v^4}-\frac{(1-\si^2)}{2} & 
\earr\]
with
\[ \de=\frac{d+x_1^2}{b_0}\qq\qq \rho=\frac{2\,x_1}{\sqrt{b_0}}-\si\qq\qq \si\in(-1,+1).\]
The parameter $m$ must vanish and this allows the change $n\to m$. Transforming the 
observables given by Theorem 3 (up to various scalings and leaving aside a reducible 
piece proportional to the hamiltonian) one obtains (\ref{firstH}) and (\ref{firstQ}) 
up to the substitutions $\Psi\to F$ and $\Ga\to G$.

For the global aspects, let us define the coordinate $t\in\,(0,1)$ by
\beq
t\equiv \tanh\frac{\chi}{2}=\frac v{1+\sqrt{v^2+1}}\,e^{\eta(v)}\qq 
\eta(v)=-\int_v^{+\nf}\left(\frac 1{\sqrt[4]{F(x)}}-\frac 1{\sqrt{1+x^2}}\right)\frac{dx}{x}
\eeq
giving for the metric
\beq
g=\Om^2\,g(S^2,{\rm can})\qq\qq  \Om^2=\frac{v^2\,B}{\sqrt{F}\,\sinh^2\chi}
\eeq
and for the hamiltonian
\beq
2H=\frac 1{\Om^2}\Big(M_1^2+M_2^2-M_3^2\Big)+\ka\,\La_1\,(\eta_1)^2+\La_2
\eeq
with
\[\qq 
\La_1=2\frac{v^2\,\sqrt{F}\,G_{,v^2}}{\sinh^2\chi}\qq\qq  
\La_2=(-\ka\,\sqrt{F}\,G_{,v^2}+l\,G+m)v^2.\]
All the functions involved do depend on $v^2$ hence on $t^2$ which is globally defined. 
They are $C^{\nf}$ for $t^2\in(0,1)$. To extend this result to $[0,1)$ we just need to check their 
behaviour for $t\to 0+$ using 
\[\frac v2=\tau+\si\,\tau^3+\tau^5+O(\tau^7)\qq\qq\qq \tau=e^{-\eta(0)}\,t.\]
The relation 
\[{\cal M}=\frac{2t}{v}\frac{F^{1/4}}{1-t^2}\left\{M_1
+\frac{(1-t^2)^2}{4t^2}\left(\frac B{\rho\,F^{1/4}}
-\frac{1+t^2}{1-t^2}\right)\,\eta_1\,M_3\right\}\]
allows to check that ${\cal M}$ is globally defined, and writing the quartic integral as
\[Q=\frac 14\,M_3^4+H\,M_3^2+\ka\frac{\rho}{B}{\cal M}^2+\La_3\,\eta_1^2\,M_3^2+\La_4\,M_3^2
+\La_5\,(\eta_1\,\eta_2)^2+\La_6\,\eta_2^2\]
where the $\La_i$ are easily retrieved from (\ref{firstQ}). Elementary checks  
for $t\to 0$ show that it is indeed globally defined. 

To discuss the sign of $B$ it is convenient to use, instead of $v\in\,[0,+\nf)$, 
the new variable $w$ defined by
\[{\rm cn}(2w)=\frac{1-v^2}{1+v^2}\qq\qq \si=1-2k^2\qq\qq w \in\,[0,K).\]
Then we have
\[B-\rho=\de\,\frac{{\rm sn}^2\,w\,{\rm dn}^2\,w}{{\rm cn}^2\,w}
-2k^2k'^2\,\frac{{\rm sn}^2\,w}{{\rm dn}^2\,w}  \qq \de\neq 0.\]
Defining $x$ for new variable
\[x=x_0\,{\rm cn}^2\,w\qq\qq x_0=\frac{k^2}{k'^2}\qq\qq   x\in\,(0,x_0]\]
we can write
\[B-\rho=\de\,k'^2\,\frac{(x_0-x)(x-r_-)(x-r_+)}{x(x+1)}\qq\qq  r_{\pm}=\frac{1-\de\pm\sqrt{1-2\de}}{\de}\]
which explicitly exhibits its three zeros.

For $\de\geq 1/2$ the roots $\,r_{\pm}$ are complex hence 
$B-\rho\geq 0$ so that if $\rho>0$ then $B>0$ for all $x\in\,(0,x_0]$. 

For $\de<0$ the roots  $\,r_{\pm}$ are both negative in which case $B-\rho\leq 0$ will imply that for $\rho<0$ we have  $B>0$ for all $x\in\,(0,x_0]$. 

The last case is $\,\de\in\,(0,1/2)$ for which we have 
$\,0<r_-<r_+$. If $r_-\geq x_0$ then $\,B-\rho\geq 0$ and then $\rho>0$ implies again that $B>0$ 
for all $x\in\,(0,x_0]$. In the remaining cases either $r_-<x_0\leq r_+$ (then $B'(x)$ has a single simple root $\,0<x_1<x_0$) or $r_-<r_+<x_0$ (then $B'(x)$ has two simple roots with 
$0<x_1<x_2<x_0$), the minimum value of $B-\rho$ is always given by $B(x_1)-\rho$ where $x_1$ is the smallest positive root of $B'(x)$ and if $\rho>B(x_1)$ then $B>0$ for all $x\in\,(0,x_0]$. 

Taking into account the relation
\[D_x\,B=-\frac{\de\,k'^2}{(x+1)^2}\,R(x)\qq   R(x)=x^2+2x-(x_0+1)(r_++r_-+1)+\frac{2x_0}{x}+\frac{x_0}{x^2}\]
we see that, in practice, to get $x_1$ one must solve for the smallest positive root of the 
quartic equation $R(x)=0$. $\quad\Box$

\vspace{5mm}
\nin{\bf Proof of (ii):} 

\nin Let us start from the hamiltonian given in (\ref{firstH}) with $\de=0$ and in which 
we change $v\,\to\,u$. It is then possible to introduce as a new variable $v$ defined by
\[{\rm cn}\,(2v)=\frac{1-u^2}{1+u^2}\qq\qq \si=k'^2-k^2\qq v\in\,(0,K).\]
The metric
\[g=B\left(dv^2+\frac{S^2}{\mu}\,d\Phi^2\right)\qq B=\rho-2k^2k'^2\frac{S^2}{D^2}\qq  \mu=1-k^2\frac{S^2\,C^2}{D^2}\]
is now multiplied by a {\em finite} conformal factor $B$.

Since $\Phi$ is azimuthal we can extend $v$ to $\,(0,2K)$. The points $v=0$ and $v=2K$ 
are apparent singularities and the manifold is $M={\mb S}^2$. Computing $G$ leads to the 
hamiltonian given in (\ref{Hdelta0}). Since the potentials
\[l\,u^2G+n\,u^2=2l\,k^2k'^2\,\frac{(k^2\,D^2-k'^2)S^4}{C^2D^2}+n\,\frac{S^2\,D^2}{C^2}\]
are singular at the ``equator" $v=K$ we must set $l=n=0$. Similar transformations in $Q$ given by (\ref{firstQ}) lead eventually to (\ref{Qdelta0}).

To ascertain the global structure we will define
\[t\equiv \tan\frac{\tht}{2}=\frac{{\rm sn}\,v}{{\rm cn}\,v+{\rm dn}\,v}\,e^{\eta(v)}\qq\qq
\eta(v)=\int_0^v\frac{\sqrt{\mu(x)}-1}{{\rm sn}\,x}\,dx\]
which gives for the metric
\[g=\Om^2\,g(S^2,{\rm can})\qq\qq \Om^2=\frac B{\mu}\frac{(1+t^2)^2\,S^2}{4t^2}.\]
The hamiltonian becomes
\[2H=\frac 1{\Om^2}(L_1^2+L_2^2+L_3^2)+\La_1\,(\ze_2^2-\ze_1^2)+\frac mB\]
with
\[\La_1=2\ka\,k^2k'^2\frac{(D^2-2k'^2)}{D^4}\frac{(1+t^2)^2\,S^2}{4t^2}.\]
Obviously the functions $\Om^2,\,\La_1$ and $B$ are $C^{\nf}$ functions of $t^2\in\,(0,1]$ and 
using the series expansion
\beq\label{inv2}
v=2\tau+\frac 23\,(2k^2-1)\,\tau^3+O(\tau^5)\qq\quad \tau=e^{-\eta(K)}\frac t{k'}\eeq
this remains true for $t^2\in\,[0,1]$, i. e. around the north pole ($\ze_3=+1$). There is no need to check for the south pole: $t^2\to +\nf$ or $\ze_3=-1$ due to the relations
\beq
t(2K-v)=\frac 1{t(v)}\qq\Longleftrightarrow\qq v(1/t)=2K-v(t).\eeq

The relation 
\[{\cal L}=\frac{2t\,\sqrt{\mu}}{S(1+t^2)}
\left\{L_1+\frac 1{4t^2}\left[\frac{BC}{\rho\,D}(1+t^2)-(1-t^2)\right]\ze_2\,L_3\right\}\]
allows to check that ${\cal L}$ is globally defined, and writing the quartic integral as
\[Q=\frac 14\,L_3^2+H\,L_3^2+\frac{\ka\rho}{B}\,{\cal L}^2+\La_2\,\ze_2^2\,L_3^2+\La_3\,L_3^2
+\La_4\,\ze_1^2\,\ze_2^2+\La_5\,\ze_2^2\]
where the $\La_i$ are easily retrieved from (\ref{Qdelta0}), elementary computations 
for $t^2\to 0$ show that it is indeed globally defined. 

Imposing the non-vanishing of the conformal factor $\,B\,$ for $v\in[0,2K]$ gives 
\[\Big(\ B>0 \quad\Longleftrightarrow\quad \rho>0\ \Big)\qq\qq 
\Big(\ B<0 \quad\Longleftrightarrow\quad \rho<2k^2\Big)\]
after some elementary computations.$\quad\Box$

\vspace{5mm}
\nin{\bf Proof of (iii):} 

\nin F has a threefold zero at $x=x_1$, so we will write $F=(x-x_1)^3(x-x_1+b)$ with $\,b>0$. Since  
$\,x\in\,(x_1,+\nf)$ it is convenient to define
\[s\equiv \sinh v=\sqrt{\frac b{x-x_1}}\qq\qq c\equiv\cosh v\qq 
v\in\,(0,+\nf)\]
It follows that
\[\sqrt{F}=b^2\,\frac c{s^4}\qq\qq \be^2=\frac{b^2}{2r}\frac B{s^2}
\qq\qq G=\frac{b^2}{16r}\,\frac{(c+3)\,s^2}{(c+1)^3}\]
where
\[\rho=-\frac{b-4x_1}{2b}\qq\qq \de=\frac{d+x_1^2}{b^2}\qq\qq B=\rho+\de\,s^2+\frac{s^2}{2(c+1)^2}\] 
The parameter $m$ must again vanish and this allows the change $n\to m$. Transforming the 
observables given by Theorem 3 (up to various scalings) one obtains (\ref{tripleH}) and (\ref{tripleQ}). 

To study the global aspects we need again the relations (\ref{coord1})
\[t\equiv\tanh\frac{\chi}{2}=\tanh \frac v2\,\,e^{\eta(v)}\qq\qq 
\eta(v)=-\int_v^{+\nf}\frac{(\sqrt{\cosh x}-1)}{\sinh x}\,dx\]
mapping $v\in(0,+\nf)\to t\in(0,1)$. We get this time
\beq
g=\Om^2\,g(H^2,{\rm can})\qq\qq \Om^2(t^2)=B\,\frac{(1-t^2)^2\,\sinh^2 v}{4t^2\,\cosh v}
\eeq
The relations (\ref{tv1}) and (\ref{DLiii}), proved for Theorem 4, case (iii), are still valid
\[\tanh\frac v2=t\,e^{-\psi(t^2)}\qq\qq v=\tau-\frac{\tau^3}{24}+O(\tau^5)\qq \tau=2e^{-\eta(0)}\,t.\]
The hamiltonian can be written
\beq
2H=\frac 1{\Om^2(t^2)}\Big(M_1^2+M_2^2-M_3^2\Big)+\ka\,\La_1(t^2)\,\eta_1^2+\La_2(t^2)
\eeq
with
\[\La_1(t^2)=\frac{\ka}{B}\frac{(1-t^2)^2}{2t^2}\frac{s^2}{(c+1)^3}\qq 
\La_2(t^2)=\frac{s^2}{B}\Big(\frac{-\ka+l(c+3)s^2}{(c+1)^3}+m\Big)\]
and is therefore globally defined on ${\mb S}^2$ while the quartic integral
\[Q=\frac 14\,M_3^4+H\,M_3^2+2\frac{\rho\ka}{B}{\cal M}^2+\La_3(t^2)M_3^2+\ka\,\La_4(t^2)\,\eta_1^2\,M_3^2
-\ka^2\,\La_5(t^2)\,\eta_1^4+\ka\,\La_6(t^2)\,\eta_2^2\]
and
\[{\cal M}=2\frac{\sqrt{c}}{1-t^2}\frac ts\left[M_1+\frac{(1-t^2)^2}{4t^2}
\left(\frac B{\rho\sqrt{c}}-\frac{(1+t^2)}{(1-t^2)}\right)\zeta_1\,M_3\right]\]
are also globally defined. 

As a last step one has to impose the non-vanishing of $B$ for $v\in\,[0,+\nf)$. The resulting constraints on the parameters are 
\[\barr{ll}B>0: \quad &\dst \de> 0\ \wedge\ \rho>0\\[4mm] 
B<0:\quad & \dst \Big(\de\leq -\frac 18\ \wedge\ \rho<0\Big)
\ \vee\ \Big(\de\in\,(-\frac 18,\,0)\ \wedge\ \rho<|\de|(x_0+1)(3x_0-1)-\frac 12\Big)\earr\]
where the cubic
\[2|\de|\,x_0(x_0+1)^2=1\]
has $x_0$ for unique real solution. $\quad\Box$  

\vspace{5mm}
\nin{\bf Proof of (iv):} 

\nin For $\de=0$ the conformal factor, given in (iii), is bounded and for $\rho=-1/2$ it 
exhibits a strong decrease for $v\to +\nf$. The metric
\[g=-\left(\frac{dv^2}{c+1}+\frac{s^2}{c(c+1)}\,d\Phi^2\right)\]
under the change of variable
\[\tht=\arctan\Big(\sinh\frac v2\Big)\qq\qq v\in\,(0,+\nf)\quad\longrightarrow\quad \tht\in\,(0,\,\pi/2)\]
becomes
\[g=-2\Big(d\tht^2+\frac{\sin^2\tht}{1+\sin^2\tht}\,d\Phi^2\Big)\]
on which it is clear that the range of $\tht$ can be extended to $\,(0,\pi)$ giving for manifold 
$\,M={\mb S}^2$. So we recover Kovalevskaya's metric (\ref{Kovamet}), but with a different potential as can be seen from the hamiltonian (\ref{GoH}) and its quartic integral (\ref{GoQ}). In fact 
Goryachev \cite{Go} derived a  generalization of Kovalevskaya with two more parameters. His  
potential, as quoted in \cite{hs}, can be written 
\[k\,\sin\tht\,\cos\phi+\sin^2\tht\Big(B_1\,\sin(2\Phi)+B_2\,\cos(2\Phi)\Big)\]
and we only got the special case $k=B_1=0$.

Writing the full system   
\beq\label{Goglobal}\barr{l}
2H=L_1^2+L_2^2+2\,L_3^2+\ka(\ze_1^2-\ze_2^2)\\[4mm]
Q=L_3^4-2H\,L_3^2+\ka\,{\cal L}^2+\ka\Big[1+\ze_3^2(1-\ze_1^2)\Big]L_3^2
+\ka^2\,\ze_2^2\,\Big(\ze_1^2-1\Big)\,\earr\eeq
with $\ {\cal L}=L_1-\ze_1\,\ze_3\,L_3$ shows explicitly that it is globally defined.$\quad\Box$

Let us now examine the last possible case where $x_1<x<x_2$ and $F=(x-x_1)^2(x_2-x)^2$. 
Let us prove

\begin{nth} The locally integrable system defined in Theorem 3 becomes globally defined 
on $M={\mb H}^2$. Its hamiltonian becomes
\beq\label{exoticH}
2H=\frac 1B\left(P_v^2+\frac{\pf^2}{\cosh^2 v}+\ka\,\frac{\tanh v}{\cosh^2 v}\,\cos\phi
+l\,\tanh^2 v+m\,\tanh v+n\right) 
\eeq
with
\[B(v)=\rho+2\si\,\tanh v+\tanh^2 v\]
and its quartic integral
\beq\label{exoticQ}
\barr{l}\dst Q=\pf^4+2H\,(\pf^2-\ka\si\,\cos\phi)-\ka(\sin\phi\,P_v-\tanh  v\,\cos\phi\,\pf)\pf-l\,\pf^2\\[4mm]\dst \qq\quad   
-\frac{\ka^2}{4}\frac{\sin^2\phi}{\cosh^2 v}+\frac{\ka m}{2}\,\cos\phi
\earr\eeq
where the conformal factor $B(v)$ should not vanish for $v\in\,{\mb R}$.
\end{nth}

\vspace{5mm}
\nin{\bf Proof :} 

\nin The change of variable
\[v=\frac 12\ln\left(\frac{x-x_1}{x_2-x}\right)\qq\qq x\in\,(x_1,\,x_2)\to v\in\,{\mb R}\]
and the change of function
\[\be^2=\frac{(x_2-x_1)^2}{4r}\,B(v)\qq\qq B(v)=\rho+2\si\,\tanh v+\tanh^2 v\]
turn the metric into
\[g=\frac{B}{r}(dv^2+\cosh^2 v\,d\phi^2).\]
The variable $\phi$ is no longer an azimuthal angle. Taking $\,(v,\,\phi) \in\,{\mb R}^2$,  
and provided that $B(v)$ does not vanish for $v\in\,{\mb R}$, we get a metric on $M={\mb H}^2$ as discussed in the Appendix A. The function $G$ becomes
\[G=\frac{(x_2-x_1)^2}{4r}\,\frac 1{\cosh^2 v}\]
and up to a few scalings of the parameters, the observables $H$ and $Q$ are seen to  
be given by the formulas (\ref{exoticH}) and (\ref{exoticQ}).

Let us first observe that the following functions
\[\phi=\sinh^{-1}\frac{\eta_1}{\sqrt{1+\eta_2^2}}\qq\qq \frac 1{\cosh^2 v}=\frac 1{1+\eta_2^2} 
\qq \qq \tanh v=\frac{\eta_2}{\sqrt{1+\eta_2^2}}\]
are globally defined. So writing the hamiltonian as
\[2H=\frac 1B\left(M_1^2+M_2^2-M_3^2+\ka\,\frac{\tanh v}{\cosh^2 v}\,\cos\phi
+l\,\tanh^2 v+m\,\tanh v+n\right)\]
shows at once its global definiteness. For the quartic integral we have
\[
\barr{l}\dst Q=M_2^4+2H\,(M_2^2-\ka\si\,\cos\phi)-\ka(\sin\phi\,P_v-\tanh v
\,\cos\phi\,M_2)M_2-l\,M_2^2\\[4mm]\dst 
\qq-\frac{\ka^2}{4}\frac{\sin^2\phi}{\cosh^2 v}+\frac{\ka m}{2}\,\cos\phi\earr\]
and all terms are globally defined if we take into account the relation
\[P_v=\frac{\eta_3\,M_1+\eta_1\,M_3}{\sqrt{1+\eta_2^2}}.\]
As a last step we need the constraints on the parameters $\,(\rho,\,\si)$ that ensure the 
non-vanishing of $B$ for $v\in\,{\mb R}$. These follow from an elementary discussion giving 
\[\barr{lcl}
B>0\quad & \Longleftrightarrow & \quad\dst \Big(\ |\si|\geq 1 \ \wedge\ \rho+1>2|\si|\ \Big)\ \vee\ \Big(\ |\si|<1\ \wedge\  \rho>\si^2\ \Big)\\[4mm]
B<0\quad & \Longleftrightarrow & \quad\quad \rho+1<-2|\si|\earr\]
ending up the proof. $\quad\Box$

\section{Lower order integrals}
An important question is whether all the integrable systems considered in the previous sections admit conserved quantities of degree strictly less than four with respect to the momentum grading.  We will write the hamiltonian
\beq\label{HLO}
2H=a^2(x)\,P_x^2+b^2(x)\,\pf^2+f(x)\,\cos\phi+g(x)
\eeq
where none of the functions $\,(a,\,b,\,f)$ can be identically vanishing and were determined in  the previous sections. Let us prove:
\begin{nth} No first degree integral is possible for the hamiltonian (\ref{HLO}).
\end{nth}

\vspace{5mm}
\nin{\bf Proof:} Let us write the extra conserved linear integral as
\[R=a(x)\,A(x,\phi)\,P_x+B(x,\phi)\,\pf.\]
The constraint $\{H,R\}=0$ involves terms of degree 2 and 0 in the momenta giving
\[\pt_x A=0 \qq \pt_x\,B=-\frac{b^2}{a}\,\pt_{\phi} A \qq \pt_{\phi} B=a\,\frac{b'}{b}\,A
\qq B=\frac af\,\frac A{\sin\phi}(g'+f'\,\cos\phi).\]
Inserting the last relation in the second one we get
\beq\label{intR1}
\left(a\,\frac{f'}{f}\right)'\,\cos\phi+\left(a\,\frac{g'}{f}\right)'=-\frac{b^2}{a}\,\sin\phi\,\frac{\dot{A}(\phi)}{A(\phi)}\qq\qq \dot{A}(\phi)=D_{\phi}\,A(\phi).
\eeq
We have supposed that $A(\phi)\not\equiv 0$ since then $B$ hence $R_1$ would vanish identically. 
The discussion has to consider two cases since $\dot{A}(\phi)$ may be identically vanishing or not.  
This leads to the constraints
\beq\label{sep}
(c_1):\qq \left(a\,\frac{g'}{f}\right)'=0\qq\qq (c_2):\qq \frac a{b^2}\,\left(a\,\frac{f'}{f}\right)'=\la\qq  
\la\in{\mb R}.
\eeq
For the models in Theorems 2 and 3 we have
\[a\,\frac{g'}{f}=\beta\left(l+\frac m{G'}\right)\]
which is a constant only for the globally defined systems of Theorem 4, for which $m=0$ and $\be=1$. But for these cases we have checked that ($c_2$) does not hold. In all the remaining cases for 
which $\be$ is not a constant, the condition ($c_1$) will not hold. $\quad\Box$

\begin{nth} No  second or third degree integral is possible for the hamiltonian (\ref{HLO}).
\end{nth}

\vspace{5mm}
\nin{\bf Proof:} Let us write the quadratic extra conserved quadratic integral as
\[R_2=a^2(x)\,A(x,\phi)\,P_x^2+a(x)\,B(x,\phi)\,P_x\,\pf+C(x,\phi)\,\pf^2.\]
Let us first notice that there is no loss of generality if we do not include a linear term 
for the following reason: when one expands
\[\{H,R_2+R_1\}=\{H,R_2\}+\{H,R_1\}\]
the first piece has linear and cubic terms which are decoupled from the quadratic and zero degree 
terms coming from the second piece, which were already shown in Lemma 3 to produce no linear integral. Such a useful observation is also valid also for higher degrees. 

The constraint $\{H,R_2\}=0$ gives six equations from which we select the following ones
\[\pt_x A=0 \qq \pt_x\,B=-\frac{b^2}{a}\,\pt_{\phi} A \qq 
B=2\,\frac af\,\frac A{\sin\phi}(g'+f'\,\cos\phi).\]
Inserting the last relation in the second one, we get an equation discussed as in Lemma 3, 
and leading to the same constraints given by (\ref{sep}) which never hold simultaneously. 

There remains to consider the case of a cubic integral
\[R_3=a^3(x)\,A(x,\phi)\,P_x^3+a^2(x)\,B(x,\phi)\,P_x^2\,\pf+a(x)\,C(x,\phi)\,P_x\,\pf^2
+D(x,\phi)\,\pf^3.\]
From $\{H,R_3\}=0$ we select the relations
\[\pt_x A=0 \qq\qq \pt_x\,B=-\frac{b^2}{a}\,\pt_{\phi} A \qq\qq  
B=3\,\frac af\,\frac A{\sin\phi}(g'+f'\,\cos\phi)\]
leading again to the constraints (\ref{sep}) and concluding the proof. $\quad\Box$

\section{Conclusion}
Following the ideas put forward by Selivanova \cite{Se1}, \cite{Se2}, \cite{hs} a 
large class of explicit integrable systems with cubic or quartic first integrals has now 
been obtained with full control of their global structure. However, let us observe that 
the {\em local} structure of a larger class of systems with quartic integrals was obtained by 
Yehia \cite{Ye}: they models exhibit a more 
general potential in the hamiltonian and many parameters, but their  
global structure remains unknown. The approach followed in all of these references relied 
on a partial differential equation first derived by \cite{Ha} and which is appropriate 
for cubic and quartic integrals but does not seem to generalize to higher degree integrals. 
However from the pioneering work of Kiyohara \cite{Ki} we know that integrals 
of any degree do exist for two-dimensional manifolds. In \cite{Va} and in this work 
a more direct analysis of the differential system leading to integrability revealed to be also  successful: it will be interesting to see in the future if, following such a path, one can 
obtain new explicit examples exhibiting integrals of degree strictly larger than four.

\begin{appendix}
\section*{Appendix A: notational conventions}
The riemannian metrics dealt with in this article have the generic form
\beq
g=A^2(v)\,dv^2+B^2(v)\,d\phi^2\eeq
Our definitions of the scalar curvature and of the Ricci tensor are 
\[R=-\frac 2{AB}\left(\frac{B'}{A}\right)'\qq\qq Ric_{ij}=\frac R2\,\,g_{ij}.\]

Let us first consider ${\mb R}^3$ equipped with the coordinates $(\,\ze_1,\,\ze_2,\,\ze_3\,)$. 
Then the two-dimensional sphere $\ze_1^2+\ze_2^2+\ze_3^2=1$ is embedded according to
\[\ze_1=\sin\tht\,\cos\phi\qq \ze_2=\sin\tht\,\sin\phi\qq \ze_3=\cos\tht\qq \tht\in\,(0,\pi)\]
while the angle $\phi$ is azimuthal, which means that $\phi\in\,[0,2\pi)$ is extended to $[0,2\pi]$  upon identification of $\phi=0$ and $\phi=2\pi$. The induced metric is
\beq\label{canS}
g(S^2,{\rm can})=d\ze_1^2+d\ze_2^2+d\ze_3^2=d\tht^2+\sin^2\tht\,d\phi^2\qq\qq R=2.
\eeq
Let us notice that the points $\tht=0$ and $\tht=\pi$ are just apparent coordinate singularities 
(the ``poles") which must be added to get the full manifold $\,{\mb S}^2$.

We will take as generators of $so(3)$ in $T^*{\mb S}^2$
\[L_1=\sin\phi\,\pth+\frac{\cos\phi}{\tan\tht}\,\pf\qq  L_2=-\cos\phi\,\pth+\frac{\sin\phi}{\tan\tht}\,\pf\qq L_3=\pf\]
with the Poisson brackets
\[\{L_1,L_2\}=-L_3\qq\qq \{L_2,L_3\}=-L_1\qq\qq \{L_3,L_1\}=-L_2\]
which allow to write the hamiltonian 
\[2H=\pth^2+\frac{\pf^2}{\sin^2\tht}=L_1^2+L_2^2+L_3^2.\]

Similarly  ${\mb H}^2$ which is defined, taking $\,(\eta_1,\,\eta_2,\,\eta_3)$ for the coordinates 
in ${\mb R}^3$, by
\[\eta_1^2+\eta_2^2-\eta_3^2=-1\qq (\eta_1,\eta_2)\in{\mb R}^2\qq \eta_3\geq 1\] 
will be embedded according to
\[\eta_1=\sinh\chi\,\cos\phi\qq \eta_2=\sinh\chi\,\sin\phi\qq \eta_3=\cosh\chi\qq\chi\in\,(0,+\nf)\]
where $\phi$ is again azimuthal. The induced metric is 
\beq\label{canH1}
g(H^2,{\rm can})=d\eta_1^2+d\eta_2^2-d\eta_3^2=d\chi^2+\sinh^2\chi\,d\phi^2\qq\qq R=-2.
\eeq
This time the point $\chi=0$ is again an apparent coordinate singularity, which must be added 
to get the manifold $\,{\mb H}^2$, while at infinity the metric takes the 
characteristic form
\[g\sim du^2+e^u\,d\phi^2\qq\qq u=2\chi\to +\nf\]

We will take as generators of $so(2,1)$ in $T^*{\mb H}^2$
\[M_1=\sin\phi\,P_{\chi}+\frac{\cos\phi}{\tanh\chi}\,\pf\qq  M_2=-\cos\phi\,P_{\chi}+\frac{\sin\phi}{\tanh\chi}\,\pf\qq M_3=\pf\]
with the Poisson brackets
\[\{M_1,M_2\}=M_3\qq\qq \{M_2,M_3\}=-M_1\qq\qq \{M_3,M_1\}=-M_2\]
which give for hamiltonian 
\[2H=P_{\chi}^2+\frac{\pf^2}{\sinh^2\chi}=M_1^2+M_2^2-M_3^2.\]

As is well known, the embedding of ${\mb H}^2$ in ${\mb R}^3$ is not unique and as 
we will experience  in Theorem 10 the following embedding
\[\eta_1=\cosh v\,\sinh\phi\qq \eta_2=\sinh v\qq \eta_3=\cosh v\,\cosh\phi\qq 
(v,\,\phi)\in\,{\mb R}^2\]
is useful. The induced metric is 
\beq\label{canH2}
g(H^2,{\rm can})=d\eta_1^2+d\eta_2^2-d\eta_3^2=dv^2+\cosh^2 v\,d\phi^2\qq\qq R=-2
\eeq
and the generators of $so(2,1)$ in $T^*{\mb H}^2$ become
\[M_1=\cosh\phi\,P_v-\tanh v\,\sinh\phi\,\pf\qq M_2=\pf\qq 
M_3=-\sinh\phi\,P_v+\tanh v\,\cosh\phi\,\pf\]
giving for hamiltonian 
\[2H=P_v^2+\frac{\pf^2}{\cosh^2 v}=M_1^2+M_2^2-M_3^2.\]

\end{appendix}

\end{document}